\documentclass{iopart}

	\usepackage{iopams}
	\usepackage{setstack}
	\usepackage[utf8]{inputenc}	
	\usepackage{graphicx}
	\usepackage{subfigure}
	\usepackage{cite}
	\usepackage{hyperref}
	\usepackage{color}

\begin{document}

\title{Relativistic Gravitational Phase Transitions and Instabilities of the Fermi Gas}
\author{Zacharias Roupas}
\address{Department of Mathematics, University of Ioannina, Ioannina 45110, Greece} 
\author{Pierre-Henri Chavanis}
\address{Laboratoire de Physique Th\'eorique, Universit\'e Paul Sabatier, 118 route de Narbonne 31062 Toulouse, France} 


\begin{abstract}
We describe microcanonical phase transitions and instabilities of 
the ideal Fermi gas in general relativity at nonzero temperature confined in the
interior of a 
spherical shell. The thermodynamic behaviour is governed by the compactness of
rest mass, namely of the total rest mass over radius of the system. For a fixed
value of rest mass compactness, we study the caloric curves as a function of the
size of the spherical box. 
At low compactness values, low energies and for sufficiently big systems 
the system is subject to a gravothermal catastrophe, which
cannot be halted by quantum degeneracy pressure, and the system collapses. 
For small systems, there appears no instability at low energies.
For intermediate sizes, between two marginal values, gravothermal catastrophe is halted and a microcanonical phase transition occurs from a gaseous phase to a condensed phase with a nearly degenerate core. The system is subject to a relativistic instability at low energy, when the core gets sufficiently condensed above the Oppenheimer-Volkoff limit. For sufficiently high values 
of rest mass compactness the microcanonical phase transitions are suppressed.
They are replaced either by an Antonov type gravothermal catastrophe for
sufficiently big systems or by stable equilibria for small
systems. At high energies the system is subject to the `relativistic
gravothermal instability', identified by Roupas in \cite{roupas}, for all values
of compactness and any size. 
\end{abstract}

\maketitle

\section{Introduction}

The stability of self-gravitating systems in the framework of statistical mechanics was 
for the first time studied by Antonov \cite{antonov} in the case of nonrelativistic classical particles like stars in globular clusters. He considered the problem of maximizing the
Boltzmann entropy at fixed mass and energy (he had to enclose the particles
within a spherical box of radius $R$ in order to prevent the evaporation of the
system). He showed that the Boltzmann entropy has no global
maximum but that it may have a local maximum, corresponding to a
star system with a Maxwell-Boltzmann distribution, provided
that the density contrast between the center and the boundary of the box is less
than $709$. Lynden-Bell and Wood \cite{lbw} confirmed and extended the results of Antonov
\cite{antonov} by calculating the series of equilibria of
self-gravitating isothermal spheres using the results  known
in the context of stellar structure \cite{emden,chandrabook}. Indeed, the
equation of state of a stellar system with a Maxwell-Boltzmann distribution is that of an
isothermal gas with an equation of state  $P=\rho k_B T/m$ (here $\rho$ is the
mass density). They showed that the
caloric curve $T(E)$ forms a spiral\footnote{This caloric curve was first plotted (by hand) by
Katz \cite{katzpoincare1}.} and that no equilibrium
state exists in the microcanonical ensemble below a minimum energy $E_{\rm min}=-0.335GM^2/R$.
Similarly, there is no equilibrium state in the canonical ensemble below a minimum
temperature  $T_{\rm min}=GMm/(2.52Rk_B)$ (this result was already known to 
Emden; see Chapter XI of \cite{emden}). They studied
the thermodynamic stability of isothermal spheres by using the
Poincar\'e theory of linear series of thermodynamic equilibria. They showed that the
instability in the microcanonical ensemble occurs at the first turning point of
energy $E_{\rm min}$, corresponding to a density contrast of $709$, in agreement with
the result of Antonov \cite{antonov}. Similarly, in the canonical ensemble, the
instability occurs at the first turning point of temperature $T_{\rm min}$,
corresponding to a density contrast of $32.1$. They interpreted these instabilities in relation
to the negative specific heats of self-gravitating systems and introduced the term
``gravothermal catastrophe'' to name the instability discovered by Antonov.

The study of Lynden-Bell and Wood \cite{lbw} was completed by Horwitz and Katz \cite{hk3} and Katz \cite{katzpoincare1}  who generalized the
turning point criterion of Poincar\'e. They applied it to different statistical
ensembles (microcanonical, canonical and grand canonical) and
established that statistical ensembles have a different physical
meaning in long-range interacting systems, and that they are not equivalent
regarding the stability properties of thermal
equilibria.\footnote{This notion of ensemble inequivalence for
systems with long-range interactions is now well-known (see, e.g.,
\cite{campa-et-al}). This is to be contrasted to the case of systems with
short-range interactions for which the statistical ensembles are equivalent in
the thermodynamic limit \cite{hill}.}
Padmanabhan \cite{paddyapj} provided a simplification of the calculations of Antonov regarding the stability of
isothermal spheres in the microcanonical ensemble based on the sign of the
second variations of entropy. Chavanis \cite{aaiso,grand} adapted the method of
Padmanabhan  \cite{paddyapj} to the canonical ensemble \cite{aaiso} and to other
ensembles \cite{grand}, thereby recovering and extending the results of Lynden-Bell and Wood \cite{lbw},
Horwitz and Katz \cite{hk3}, and Katz \cite{katzpoincare1}. The same results
were obtained from a field theory approach by de Vega and Sanchez
\cite{dvs1,dvs2}. Some reviews on the subject are given in \cite{paddy,katzrevue,ijmpb}. 

Sorkin {\it et al.} \cite{sorkin} and, more
recently, Chavanis \cite{aarelat1,aarelat2} have considered the statistical mechanics of a self-gravitating radiation confined within a cavity in general relativity. Radiation is equivalent to a
relativistic gas of massless bosons (photons) with a linear equation of state
$P=\epsilon/3$, where $\epsilon$ denotes the energy density (this equation of
state also corresponds to the ultra-relativistic limit of an ideal gas of any kind of massive
particles, classical, fermions or bosons).
They showed that the caloric curve $T_{\infty}(E)$ (here $T_{\infty}$ denotes the temperature at
infinity) forms a spiral
and that no equilibrium state exists
above a maximum energy $E_{\rm max}=0.246\, Rc^4/G$ for an isolated system or above a maximum
temperature  $k_B(T_\infty)_{\rm max}=0.445\,(\hbar^3c^7/GR^2)^{1/4}$ for a 
system in contact with a heat bath (see Fig. 15 of
\cite{aarelat2}).
The system becomes unstable when it is too ``hot'' because energy is mass so it gravitates.  It can be shown \cite{sorkin,aarelat1,aarelat2,Roupas_2013CQG} 
that the series of equilibria becomes dynamically and thermodynamically unstable
after the first turning point of
energy, in agreement with the  Poincar\'e criterion. This corresponds to a
density contrast $22.4$ \cite{aarelat1,aarelat2}. Gravitational
collapse is expected to lead to the formation of a black
hole.

The statistical mechanics of relativistic classical self-gravitating systems was
studied by Roupas \cite{roupas,Roupas_RGI_2018}, who found that the caloric curve has the form of
a double spiral. He identified an instability of the ideal gas at high
energies, the \textit{high-energy gravothermal instability} caused by the gravitation of thermal energy. At low energies he showed that a relativistic generalization of gravothermal catastrophe, the `low-energy gravothermal instability', sets in. The double spiral reflects the two types of a gravothermal instability and shrinks as the compactness $2GmN/Rc^2$ approaches the critical value $0.3528$. Above this value no equilibrium is achievable under any conditions.

The nonrelativistic self-gravitating fermions were studied by Hertel and Thirring \cite{ht} and Bilic and Viollier \cite{bvn}. Again, it is necessary to confine the system within a box in order to prevent
its evaporation. They generalized at nonzero temperatures the results obtained
at $T=0$ by Fowler \cite{fowler}, Stoner \cite{stoner29}, Milne \cite{milne} and
Chandrasekhar \cite{chandra31nr} in the context of white dwarfs. In the
canonical ensemble they evidenced
a first order phase transition below a critical temperature from a gaseous phase to a condensed phase (fermion star). This
canonical phase transition bridges a region of negative specific heats in the
microcanonical ensemble. This phase transition occurs  provided the size
of the system is sufficiently large (for a given number of particles).  A more
general study was made by Chavanis \cite{ijmpb} who found that the
self-gravitating Fermi gas exhibits two critical points, one in each
ensemble. Small systems with $R<R_{\rm CCP}(N)$ do not experience any phase
transition, intermediate
size systems with $R_{\rm CCP}(N)<R<R_{\rm MCP}(N)$ experience a canonical phase
transition and large systems with $R>R_{\rm MCP}(N)$ 
experience both canonical and microcanonical phase transitions. When quantum
mechanics is taken into account for nonrelativistic systems, an equilibrium
state exists for any value of energy and temperature. In other words,
the pressure arising from the Pauli exclusion principle is able to prevent the
gravitational collapse of nonrelativistic classical isothermal spheres.  

Oppenheimer \& Volkoff \cite{ov} studied the statistical mechanics of self-gravitating ideal gas of fermions in general relativity at zero temperature and identified a relativistic instability for sufficiently high masses. They determined this maximum mass $M_{OV} = 0.7M_\odot$ for ideal neutron cores.
Roupas \cite{Roupas_2015PRD} generalized to all temperatures the original
calculation of Oppenheimer \& Volkoff, providing the analogue of Oppenheimer-Volkoff analysis for the whole cooling stage of a neutron star; from the ultra hot progenitor, the proto-neutron star \cite{Burrows:1986me,Prakash:1997,Lattimer_2016PhR...621..127L}, down to the final cold star. 
Bilic and Viollier \cite{bvr}, earlier, had studied the statistical mechanics of self-gravitating fermions in general relativity confined in a box. They considered specific values of parameters for one particular
situation where the number of particles $N$ is below the `Oppenheimer-Volkoff
(OV) limit' $N_{\rm OV}$, namely the maximum $N$ at zero temperature and zero
boundary pressure for which the system is stable (see 
\ref{sec_ov})
and the radius of the system is large enough
so that a first order canonical phase transition from a gaseous phase to a condensed phase occurs like in the nonrelativistic case. A more general
study of phase transitions in the  general relativistic Fermi gas was made by
Alberti and Chavanis \cite{ac2} who determined the 
complete phase diagram of the system in the
$(R,N)$ plane. They showed in particular that for a fixed radius there is no equilibrium state
below a critical temperature or below a critical energy when $N>N_{\rm OV}$. 
In that case, the system is expected to collapse since quantum
degeneracy pressure cannot stabilize the system anymore. Alberti and Chavanis \cite{ac2}
 studied the caloric curves and the phase transitions in the general
relativistic Fermi gas by fixing the system size $R$ and
varying the number of particles $N$. In this paper, we shall use the system
size $R$ and the \textit{compactness} of rest mass
$GNm/Rc^2$ as independent variables. The classical limit is recovered for
$R,N\rightarrow +\infty$ with $N/R$ fixed. This approach will allow us to study
quantum corrections to the classical limit when $R$ is reduced.

In the next section we review the relativistic Fermi gas. In section \ref{sec:TOV} we setup the problem in the general relativistic context and define our control parameters. In section \ref{sec:micro} we identify the gravitational phase transitions and instabilities and present our main results. We discuss our conclusions in section \ref{sec:conclusions}. In the  \ref{app:sec_pd} we discuss the various regimes of our control parameters with respect to the results of \cite{ac2}.

\section{The relativistic Fermi gas}

For an ideal relativistic quantum gas \cite{landsberg}, the
one-particle energy distribution is given by the
Fermi-Dirac or Bose-Einstein distributions for fermions or bosons respectively:
\begin{equation}\label{eq:quantum_dis}
	g(\epsilon) = \frac{1}{e^{\beta(\epsilon - \mu)}\pm 1} \;,\;
	\left\lbrace
	\begin{array}{l}
		(+)\;\mbox{for fermions} \\[2ex]
		(-)\;\mbox{for bosons}
	\end{array}
	\right.
\end{equation}
where $\epsilon$ is the energy per particle, including rest mass in the relativistic case, $\mu$ the chemical potential and $\beta = 1/kT$ the inverse temperature. Substituting the relativistic definition of energy 
\begin{equation}\label{eq:e_p}
	\epsilon = \sqrt{m^2c^4 + p^2 c^2},
\end{equation}
where $m$ is the mass of one particle and $p$ its momentum, and applying the 
Juttner transformation 
\begin{equation}\label{eq:Juettner}
	\frac{p}{mc} = \sinh\theta,
\end{equation}
the distribution \eref{eq:quantum_dis} may be written in terms of $\theta$ as
\begin{equation}\label{eq:fb_dis}
	g(\theta) = \frac{1}{e^{b \cosh\theta - \alpha}\pm 1}, 
\end{equation}
where 
\begin{equation}
	b = \frac{mc^2}{kT }
\end{equation}
and 
\begin{equation}
	\alpha = \frac{\mu}{kT}. 
\end{equation}

Let us focus on the case of fermions. The phase space one-particle distribution function for quantum degeneracy $g_s$ (e.g. $g_s = 2$ for neutrons) is 
\begin{equation}\label{eq:f_dis-f}
	f(\vec{r},\vec{p}) = \frac{g_s}{h^3}g(\epsilon),
\end{equation} 
where $h$ is Planck constant. 
It is rather straightforward using the distribution \eref{eq:f_dis-f} to show
\cite{chandrabook} that the pressure $P$, number density $n$ and total
mass-energy density $\rho$ may be written as 
\begin{eqnarray}
\label{eq:P_Q} 	
&P \equiv \frac{1}{3}\int f(\vec{r},\vec{p})p\frac{\partial \epsilon}{\partial p}d^3\vec{p} =  \frac{4\pi g_s m^4 c^5}{3h^3}\int_0^\infty \frac{\sinh^4\theta
d\theta}{e^{b \cosh\theta - \alpha}+1},
\\
\label{eq:rho_Q}
&\rho \equiv \int f(\vec{r},\vec{p}) \epsilon d^3\vec{p}  = \frac{4\pi g_s m^4 c^3}{h^3}\int_0^\infty \frac{\sinh^2\theta
\cosh^2\theta  d\theta}{e^{b \cosh\theta - \alpha}+1},
\\
\label{eq:n_Q}
&n \equiv \int f(\vec{r},\vec{p})  d^3\vec{p} = \frac{4\pi g_s m^3 c^3}{h^3}\int_0^\infty \frac{\sinh^2\theta \cosh\theta  d\theta}{e^{b \cosh\theta - \alpha}+1}
,
\end{eqnarray}

Following Chandrasekhar \cite{chandrabook} we define the functions $J_\nu (\alpha,b)$ as
\begin{equation}\label{eq:Jnu}
	J_\nu (\alpha, b) = \int_0^\infty \frac{\cosh(\nu \theta)}{ e^{b \cosh\theta-\alpha} + 1 } d\theta.
\end{equation}
Equations \eref{eq:P_Q}, \eref{eq:rho_Q}, \eref{eq:n_Q} may then be written 
as 
\begin{eqnarray}
\label{eq:P_Qb}
&P = \frac{4\pi g_s m^4 c^5}{3h^3} \left( \frac{3}{8} J_0 - \frac{1}{2} J_2 +
\frac{1}{8} J_4 \right), 
\\
\label{eq:rho_Qb}
&\rho = \frac{4\pi g_s m^4 c^3}{h^3} \left( - \frac{1}{8} J_0 + \frac{1}{8} J_4
\right),
\\		
\label{eq:n_Qb}
&n = \frac{4\pi g_s m^3 c^3}{h^3} \left( - \frac{1}{4} J_1 + \frac{1}{4} J_3
\right).
\end{eqnarray}
The equation of state may be expressed with the doublet $P,\rho$ above. A formulation in different variables is achieved by use of the, so called, generalized Fermi-Dirac integrals as in \cite{Cox_1968pss}.

Let us now briefly discuss the completely degenerate
and
nondegenerate (classical) limits.  The parameter $\alpha = \mu/kT$ controls the degeneracy of the system. We have the following limits:
\begin{eqnarray}
\label{eq:limit_Q}		&\alpha \rightarrow +\infty :
\mbox{ Completely degenerate limit} \\
\label{eq:limit_C}		&\alpha \rightarrow -\infty :
\mbox{ Nondegenerate (classical) limit} 
\end{eqnarray}

We stress that the second criterion is sufficient but not necessary. The classical limit may apply for any $\alpha$, positive or negative, provided that $\beta\epsilon-\alpha\gg 1$.

In the first case (\ref{eq:limit_Q}), the chemical potential is positive and large compared to the temperature and we denote it $\mu = \epsilon_F$. The distribution function (\ref{eq:quantum_dis}) becomes:
\begin{equation}
	g(\epsilon) \overset{\alpha \rightarrow \infty}{\longrightarrow}\left\lbrace
	\begin{array}{l}
		1, \;\epsilon \leq \epsilon_F \\[2ex]
		0, \;\epsilon > \epsilon_F
	\end{array}
	\right.
\end{equation}
Thus, the integrals (\ref{eq:P_Q}-\ref{eq:rho_Q}) have an upper limit $p_F$ and we get
\begin{eqnarray}
\label{eq:P_Qd} 	
&P =  \frac{4\pi g_s m^4 c^5}{3h^3}\int_0^{p_F} \sinh^4\theta d\theta ,
\\
\label{eq:rho_Qd}
&\rho = \frac{4\pi g_s m^4 c^3}{h^3}\int_0^{p_F} \sinh^2\theta \cosh^2\theta  d\theta,
\\
\label{eq:n_Qd}
&n = \frac{4\pi g_s m^3 c^3}{h^3} \int_0^{p_F} \sinh^2\theta \cosh\theta  d\theta.
\end{eqnarray}
The integration may be performed analytically as in p. 360 of \cite{chandrabook}. The chemical potential $\mu=\epsilon_F$ is identified with the Fermi energy. 

In the second case (\ref{eq:limit_C}), the chemical potential is large and
negative $-\mu \gg kT$  leading to the Boltzmann distribution
\begin{equation}
	g(\epsilon) \overset{\alpha \rightarrow -\infty}{\longrightarrow} e^{-\beta (\epsilon - \mu)}.
\end{equation}
The integral $J_\nu (\alpha, b)$ becomes the modified Bessel function $K_\nu (b)$
\begin{equation}\label{eq:bessel}
\fl	\lim_{\alpha \rightarrow -\infty} J_\nu(\alpha,b) = e^\alpha K_\nu(b)\; ,\; 
	K_\nu(b) = \int_0^{\infty} e^{-b \cosh\theta}\cosh(\nu\theta)d\theta.
\end{equation}
Using the recursive relations
\begin{equation}\label{eq:recur}
	K_{\nu+1}(b) - K_{\nu-1}(b) = \frac{2\nu}{b}K_\nu(b),
\end{equation}
equations \eref{eq:P_Qb}, \eref{eq:rho_Qb} and \eref{eq:n_Qb} become 
\begin{eqnarray}
\label{eq:P_clasB} 	
&P = \frac{4\pi g_s m^4 c^5}{h^3}e^{\alpha}\frac{K_2}{b^2},
\\
\label{eq:rho_clasB}
&\rho = \frac{4\pi g_s m^4 c^3}{h^3}e^{\alpha}\frac{K_2}{b}(1+\mathcal{F}),
\\
\label{eq:n_clasB}
&n = \frac{4\pi g_s m^3 c^3}{h^3}e^{\alpha}\frac{K_2}{b},
\end{eqnarray}
where
\begin{equation}\label{eq:F}
		 \mathcal{F}(b) = \frac{K_1(b)}{K_2(b)} + \frac{3}{b} - 1.
\end{equation}
These give the equation of state in the classical relativistic limit
\begin{equation}\label{eq:clas_eos}
	P = \frac{n m  c^2}{b} \; \mbox{or equivalently}\;
	P = \frac{\rho c^2}{b(1+\mathcal{F})}.
\end{equation}

\section{TOV equation}\label{sec:TOV}

The Tolman-Oppenheimer-Volkoff (TOV) equation (\ref{eq:TOV}) expresses the
condition of hydrostatic equilibrium for a spherical, perfect fluid in general
relativity and may be derived from Einstein's equations (e.g.
\cite{Weinberg_1972gcpa.book}), together with equation (\ref{eq:massd}) for the
total mass-energy $\hat{M}(r)$ contained within radius $r$:
\begin{eqnarray}
	\label{eq:TOV}
\frac{dP}{dr} = -(\rho + \frac{P}{c^2})\left( \frac{G\hat{M}(r)}{r^2} + 4\pi G \frac{P}{c^2}r\right)
	\left( 1 - \frac{2G\hat{M}(r)}{rc^2} \right)^{-1},		\\
\label{eq:massd}
\frac{d\hat{M}}{dr} = 4\pi r^2 \rho.
\end{eqnarray}
We denote with $P$ the pressure and $\rho$ the total mass-energy density (rest $+$ gravitational $+$ kinetic) of the system. We reserve the symbol $M$ with no hat for the total mass-energy of the system until the boundary radius $R$ of the sphere, i.e. 
\begin{equation}
	M = \hat{M}(R) = \int_0^R \rho(r)\, 4\pi r^2 dr.
\end{equation}
The entropy is written as
\begin{equation}\label{eq:entropy}
	S = \int_0^R s(r)\left( 1 - \frac{2G\hat{M}(r)}{rc^2} \right)^{-\frac{1}{2}} 4\pi r^2 dr
\end{equation}
and the number of fermions is given by
\begin{equation}\label{eq:number}
	N = \int_0^R n(r)\left( 1 - \frac{2G\hat{M}(r)}{rc^2} \right)^{-\frac{1}{2}} 4\pi r^2 dr.
\end{equation}
where the entropy density and particle number density $s$, $n$ satisfy the Euler's relation (sometimes called integrated Gibbs-Duhem relation)
\begin{equation}\label{eq:euler}
	T s = \rho c^2 + P - \mu n.
\end{equation}
	The temperature $T$ measured by a local observer at $r$ is not constant in equilibrium in General Relativity \cite{Tolman:1930,Tolman-Ehrenfest:1930}, so that $T=T(r)$. It follows the distribution according to the differential equation
\begin{equation}\label{eq:Tprime}
	\frac{T'}{T} = \frac{P'}{P + \rho c^2}.
\end{equation}
In General Relativity, 
the thermodynamic parameter conjugate to the energy
\cite{Roupas_2013CQG,Roupas_2015CQG_32k9501R} is not the inverse of the local
temperature but the inverse of the so-called Tolman temperature. It is
constant and homogeneous at equilibrium and identified with the temperature
measured by an observer at infinity,
\begin{equation}\label{eq:Tolman_T}
	\tilde{T} = T(r) \sqrt{g_{tt}} = const.\Rightarrow
	\tilde{T} = T(R) \left( 1 - \frac{2GM}{Rc^2} \right)^{\frac{1}{2}}.
\end{equation}
 
Quantum mechanics introduces a scale to the system, 
namely the elementary phase-space cell $h^3$. Combined with general relativity,
the Planck scale is obtained. Then, the rest mass $m$ of the elementary constituent of the gas determines the Oppenheimer-Volkoff (OV) scales for all quantities as follows (see \ref{sec_ov}):
\begin{eqnarray}
\label{eq:rho_star}	\rho_\star &= \frac{4\pi g_s m^4 c^3}{h^3}, \\
\label{eq:r_star}	r_\star &= \left(\frac{4\pi
G}{c^2}\rho_\star\right)^{-\frac{1}{2}}, \\
\label{eq:M_star}	M_\star &= \frac{r_\star c^2}{G}. 
\end{eqnarray}
These scales are implied by the TOV equation (\ref{eq:TOV}) and by equations
\eref{eq:P_Q} and \eref{eq:rho_Q}.
Note that these OV scales may be written as
\begin{eqnarray}
\label{eq:rho_star_P}	\rho_\star &= 4\pi g_s \frac{m}{\lambda_C^3}, \\
\label{eq:r_star_P}	r_\star &= l_P \, \frac{m_P^2}{m^2}
\sqrt{\frac{\pi}{2g_s}},\\
\label{eq:M_star_P}	M_\star &= m_P \, \frac{m_P^2}{m^2}
\sqrt{\frac{\pi}{2g_s}},
\end{eqnarray}
where $m_P=(\hbar c/G)^{1/2}$ is the Planck mass and
$l_P=(\hbar G/c^3)^{1/2}$ is the Planck
length.
We introduce the dimensionless quantities:
\begin{equation}
	x = \frac{r}{r_\star}, \quad  \quad u = \frac{\hat{M}}{M_\star}, \quad 
\quad 
	\bar{\rho} = \frac{\rho}{\rho_\star}, \quad \quad
	\bar{P} = \frac{P}{\rho_\star c^2}.
\end{equation}
Defining $\psi(x)$ by the relation 
\begin{equation}\label{eq:b_psi}
	b(x) = b(0) e^{\psi (x)},
\end{equation}
where $b(x)=mc^2/k_BT(x)$ and
combining equation \eref{eq:Tprime} with the TOV equation \eref{eq:TOV}, we
find that equations \eref{eq:TOV}, \eref{eq:massd}, \eref{eq:P_Qb} and
\eref{eq:rho_Qb} become
\begin{eqnarray}
\label{eq:TOV_ND}  
	&\frac{d \psi (x)}{dx} = {\left(\frac{u(x)}{x^2} + \bar{P}(x) x\right) \left(1 - \frac{2u(x)}{x} \right)^{-1} }, 
\\
\label{eq:mass_d_ND}  
 	&\frac{d u(x)}{d x} = \bar{\rho}(x) x^2, 
\\
\label{eq:P_Q_ND} 	
&\bar{P}(\alpha,b(x)) = \frac{1}{24}\left( 3 J_0(\alpha,b(x)) - 4 J_2(\alpha,b(x)) + J_4 (\alpha,b(x)) \right),
\\
\label{eq:rho_Q_ND}
&\bar{\rho}(\alpha,b(x)) = \frac{1}{8}\left( - J_0(\alpha,b(x)) + J_4 (\alpha,b(x))\right).
\end{eqnarray}
This forms the system of equations that determines the thermodynamic equilibria
with initial conditions:
\begin{equation}
	\psi(0) = 0, \quad  \quad u(0) = 0, \quad \quad b(0) = b_0,
\end{equation}
for some $b_0$, whose exact value is determined by the number of particles constraint.
Equations (\ref{eq:P_Q_ND}) and (\ref{eq:rho_Q_ND}) define the
equation of state of the special relativistic Fermi gas. When they are
implemented in the context of General Relativity, they are realized as local
equations with $P = P (r)$, $\rho = \rho(r)$, and $T = T(r)$. The local
relations between $P$, $\rho$, $T$ remain the same as in special relativity,
while the global behavior, i.e. the dependence on position is dictated by
gravity.

We define the \textit{compactness} $\xi$ of rest mass $\mathcal{M} = mN$
\begin{equation}
	\label{eq:xi}
	 \xi = \frac{2G\mathcal{M}}{Rc^2}	
\end{equation}
and the dimensionless radius of the system
\begin{equation}
	\label{eq:zeta}
	\zeta = \frac{R}{r_\star},
\end{equation}
that we will use as control parameters.
Introducing also the dimensionless particle density
\begin{equation}
	\bar{n} = \frac{mn}{\rho_\star} 
\end{equation}	
the number of particles constraint may be written as
\begin{equation}\label{eq:xi_con}
	\xi = \frac{2}{\zeta}\int_0^\zeta \bar{n}
x^2\left(1-\frac{2u}{x}\right)^{-\frac{1}{2}}dx = {\rm const.}
\end{equation}
In order to generate the series of equilibria at fixed $\mathcal{M} = mN$ and
$R$, we can solve the system (\ref{eq:TOV_ND}-\ref{eq:rho_Q_ND}) for a given
$b_0$, integrating $\psi$ and $u$ in an interval $x\in
[0,\zeta]$ up to a fixed $\zeta$ each time, and calculating at each iteration
the corresponding $\alpha$ which satisfies the constraint \eref{eq:xi_con}
for a fixed $\xi$. In this manner we obtain the value of
$ER/G{\cal M}^2$ and
$\tilde\beta G {\cal M}m/R$ corresponding to
that  $b_0$. By varying $b_0$ we
can obtain the complete series of equilibria corresponding to the selected
values of $\xi$ and $\zeta$. This procedure can then be repeated for various
values of $\xi$ and $\zeta$. In the following, we shall fix the rest mass compactness $\xi$ and vary
the size $\zeta$.

We stress that the rest mass compactness, given in equation (\ref{eq:xi}), is the relativistic
parameter which controls the intensity of general relativity, with
$\xi\rightarrow 0$ being related to the nonrelativistic limit (we shall see
that this is true only for small and large radii). On the other hand,
$\zeta$, which may also be written as
\begin{equation}
	\zeta = \frac{R}{l_P}\,\frac{m^2}{m_P^2}\sqrt{\frac{2g_s}{\pi}},
\end{equation}
is the quantum parameter which controls quantum degeneracy, 
with $\zeta \rightarrow \infty$ being the classical limit. A more detailed
characterization of the nonrelativistic and classical limits is given in
\cite{ac2} and in \ref{app:sec_pd}.

\section{Phase transitions and instabilities}\label{sec:micro}

In this section we provide an illustration of microcanonical phase transitions
and
instabilities in the general relativistic Fermi gas. We refer to
\cite{ac2} and \ref{app:sec_pd} for the justification of the transition values
of $\xi$ and $\zeta$ separating the different regimes discussed below.

\begin{figure}[h!t]
\begin{center}
	\subfigure[Low energy.]{ \label{fig:B_E_F100_l}
	\includegraphics[scale = 0.3]{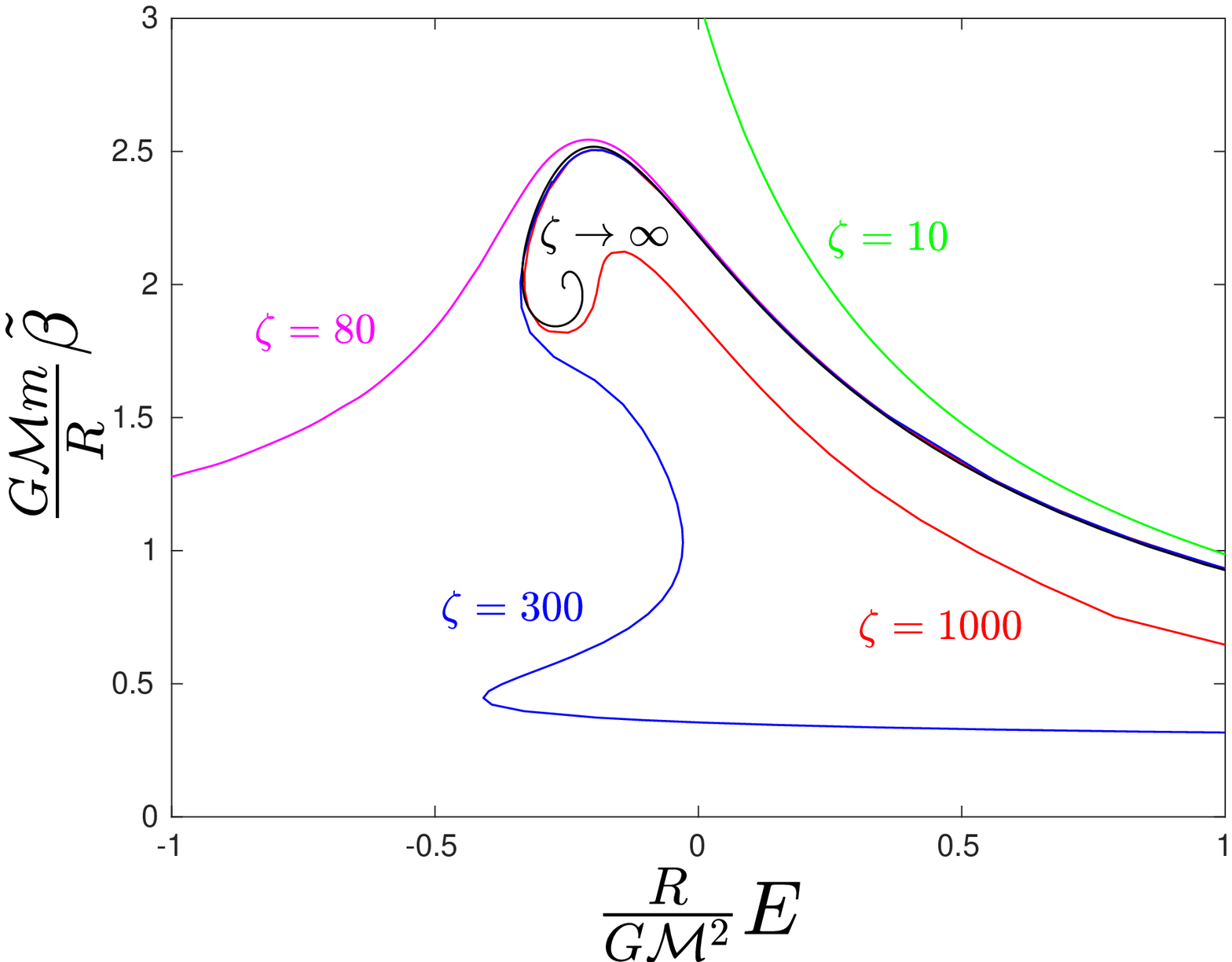} }
	\subfigure[High energy.]{ \label{fig:B_E_F100_h}
	\includegraphics[scale = 0.42]{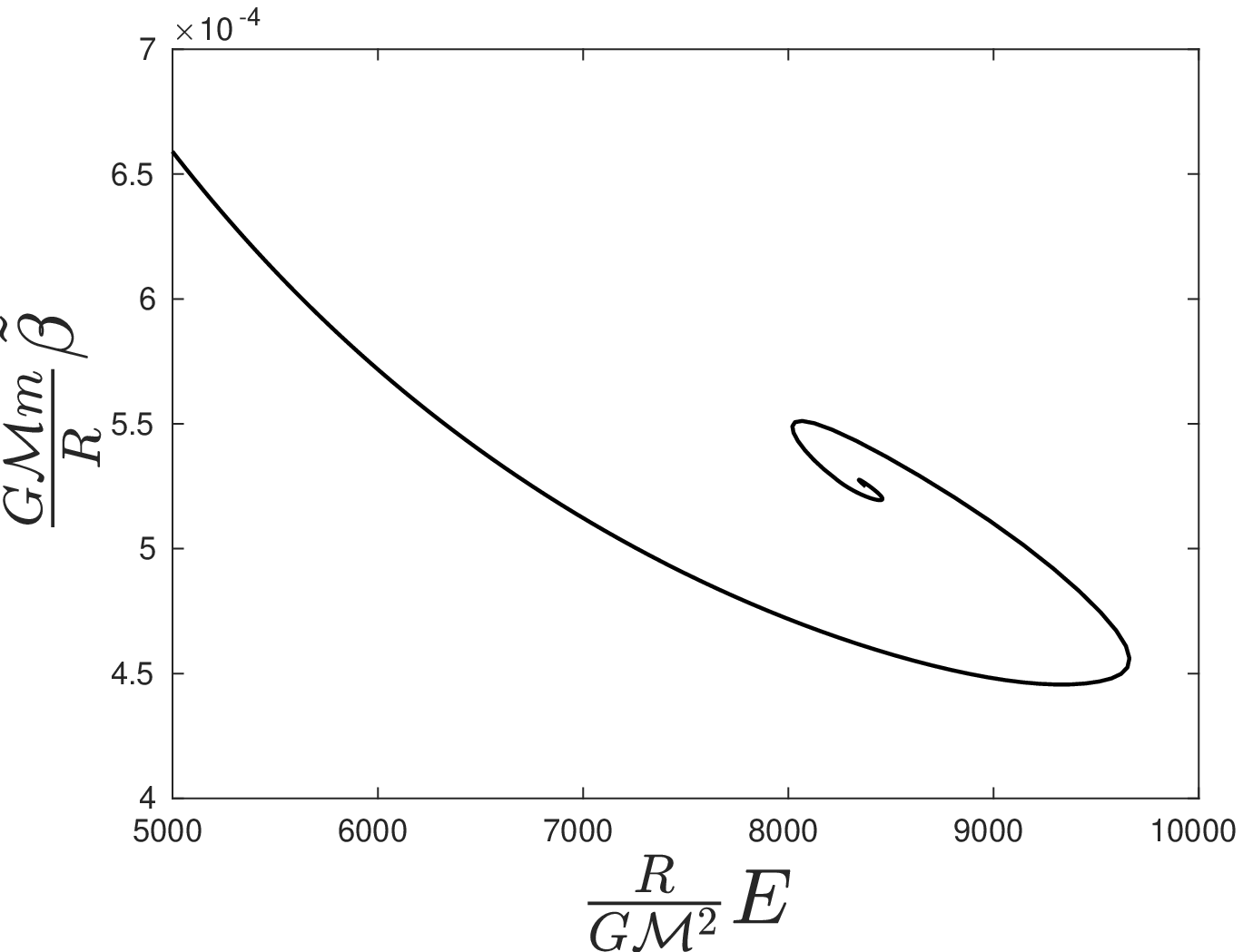} }
	\caption{Caloric curves for rest mass compactness $\xi = 0.01$ and
various values of the system size
$\zeta$. The number of particles and the size of the system are fixed. This value of $\xi$ corresponds to $R\gg
R_S$ suggesting that the system is in the Newtonian gravity limit. We denote $E$ the energy of the system subtracted by the total rest mass $E = Mc^2 - \mathcal{M}c^2$.
\textit{Left:}
The black spiral curve corresponds to the classical limit
$\zeta\rightarrow \infty$. Every curve ends to the right with an anti-clockwise
spiral, shown in the right panel, that denotes a relativistic gravothermal
instability. The curve $\zeta=80$ continues indefinitely (though not shown here) towards zero
temperature ($\tilde\beta\rightarrow +\infty$) tending asymptotically to a
minimum energy $E_{\rm min}$ like in
Figure \ref{fig:B_E_F100_R80}. The curve $\zeta=300$ designates
equilibria with a condensed
nearly degenerate core and, although the chosen value of $\xi$ implies $R\gg
R_S$, the core radius satisfies $R_{\rm C}\sim R_S$, as may be
inferred from Figure \ref{fig:MT_F100_R300_Emin}. It undergoes a
relativistic instability at sufficiently low energy when the core is
sufficiently condensed (such that $N_C>N_{\rm
OV}$).
\textit{Right:} This spiral designates the high-energy gravothermal instability \cite{roupas,Roupas_RGI_2018}. The self-gravitating gas collapses under the weight of its own thermal energy. For this value of $\xi$, the value of $\zeta$ does not alter significantly the turning point of stability, but at relativistic $\xi$ values, the $\zeta$ value does affect this instability, as shown in Figure \ref{fig:B_E_F4}.
	\label{fig:B_E_F100}}
\end{center} 
\end{figure}

In Figure \ref{fig:B_E_F100} are shown the series 
of equilibria for a compactness $\xi = 0.01$ and several values of
the system size $\zeta\ge \zeta_{\rm
min}=2.93\times 10^{-4}$. The chosen value
of
$\xi$ corresponds to $R\gg R_S$,  where
$R_S=2G{\cal M}/c^2$ is the Schwartzschild radius of the system
constructed with the rest mass ${\cal M}$. Since $R$ is much greater than $R_S$
we expect
to be in the Newtonian
gravity limit. The Newtonian gravity results are represented in
Figures 14, 21 and 31 of \cite{ijmpb}. For small systems
($\zeta<\zeta_{\rm MCP}=154$) the gravothermal
catastrophe does not occur 
and the system
passes progressively from a non-degenerate to a nearly degenerate
configuration as energy is decreasing. The caloric curve presents a
vertical asymptote at $E_{\rm min}$ corresponding to the ground state
($\tilde T=0$ or $\tilde\beta=+\infty$) of the self-gravitating Fermi
gas.\footnote{A region of negative specific
heats appears on the caloric curve when $\zeta>\zeta_{\rm CCP}=28.7$. In
the canonical ensemble, this region of negative specific heat is
replaced by a phase transition \cite{ijmpb}. On the other hand, a second
branch of equilibrium states (corresponding to unstable equilibria) appears at
$\zeta_{1}=40.8$ (see Fig. \ref{fig:B_E_F100_R80}). It presents a
vertical asymptote at $E'_{\rm min}$ corresponding to the first unstable 
state  of the self-gravitating Fermi
gas at $\tilde T=0$ (there can be up to an infinity
of  unstable states  at $\tilde T=0$). The vertical
asymptotes of the main and secondary branches merge at $\zeta_{\rm
OV}=90.0$  marking the absence of a ground state for the self-gravitating Fermi
gas beyond that point and the onset of a relativistic
instability at sufficiently low energies and temperatures.} For larger
systems ($\zeta>\zeta_{\rm MCP}=154$) the
gravothermal catastrophe does occur. In the Newtonian gravity case, the
collapse is halted by a degenerate configuration in any case \cite{ijmpb}. When
general relativity is taken into account, the system is subject to a
relativistic instability at
sufficiently low energy. The reason is
that, following the gravothermal
catastrophe, the system takes a
core-halo structure with a dense degenerate core of mass
$M_{\rm C}$ (equal to a fraction $\sim 1/3$ of the total mass $M$ \cite{ac2})
and size
$R_{\rm C}\ll R$ surrounded by an  essentially
nondegenerate isothermal halo. 
This type of core-halo configuration renders the total
size
of the system $R$ irrelevant. In that case, what
determines the validity, or the invalidity,  of the Newtonian gravity
approximation is the size of the
core. Therefore,  for large enough systems  the
Newtonian gravity approximation breaks down because $R_{\rm C}\sim R_S$ even
though $R \gg R_S$. When the degenerate core becomes sufficiently condensed, it
collapses.

\begin{figure}[htb]
\begin{center}
	\subfigure[Inverse temperature]{ \label{fig:E_F100_R300}
	\includegraphics[scale = 0.3]{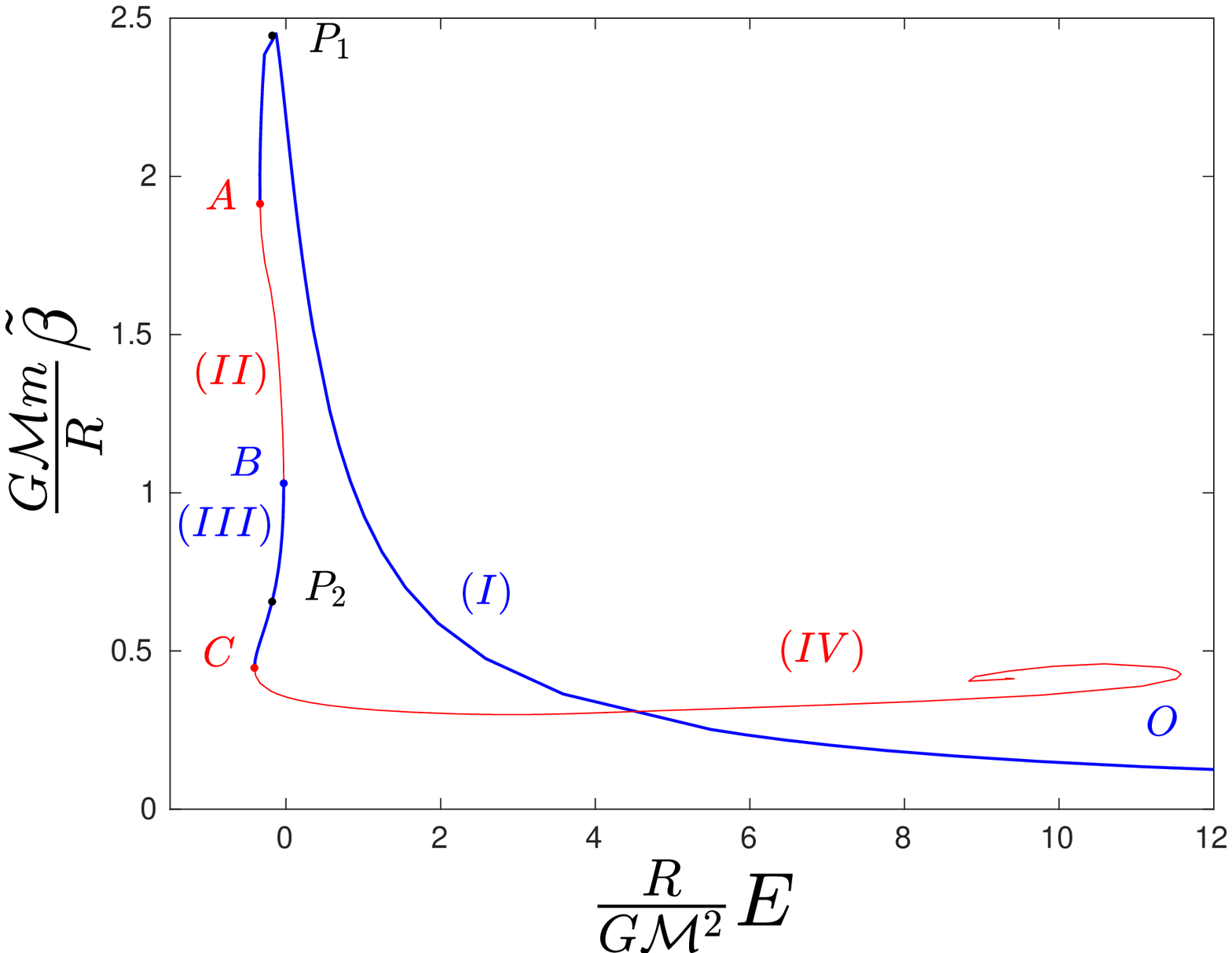} } 
	\subfigure[Entropy]{ \label{fig:S_F100_R300}
	\includegraphics[scale = 0.3]{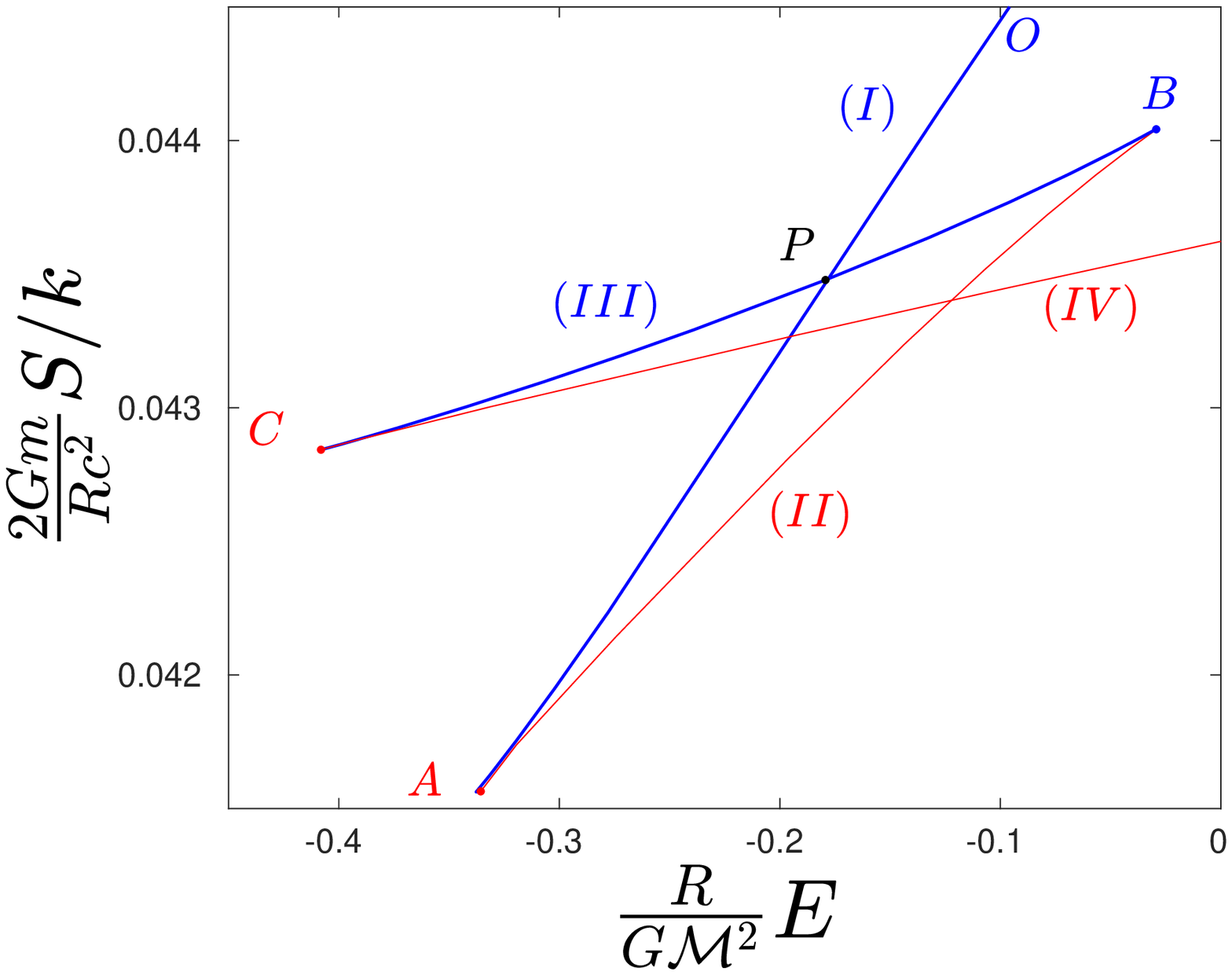} } 
	\caption{$\xi = 0.01$, $\zeta = 300$. Blue, thick branches $(I)$ and
$(III)$ are stable, while red, thin branches $(II)$ and $(IV)$ are unstable in
the microcanonical ensemble. The stability of the solutions can be deduced
from the Poincar\'e turning point criterion \cite{poincare}. Branch $(I)$ is in
the \emph{gaseous phase}
while branch $(III)$ is in the \emph{condensed phase}. Point $P$
designates a first order phase transition from configuration $P_1$ to $P_2$ (it
can be determined from the Maxwell construction or from the equality of the
entropy of the two phases). We note that the first order phase transition does
not take place in practice because of the very long lifetime of metastable
states, scaling as $e^N$, for systems with long-range interactions \cite{ijmpb}.
Therefore, the physical phase transition is the zeroth order phase transition
which takes place at the spinodal point
$A$ where the metastable gaseous phase disappears (gravothermal catastrophe). At
point $C$ a relativistic instability sets in to the nearly degenerate, condensed
core, which collapses further when the number of particles that it contains
passes above the OV limit.
	\label{fig:F100_R300}}
\end{center} 
\end{figure}

\begin{figure}[htb]
\begin{center}
	\subfigure[Density]{ \label{fig:rho_F100_R300}
	\includegraphics[scale = 0.3]{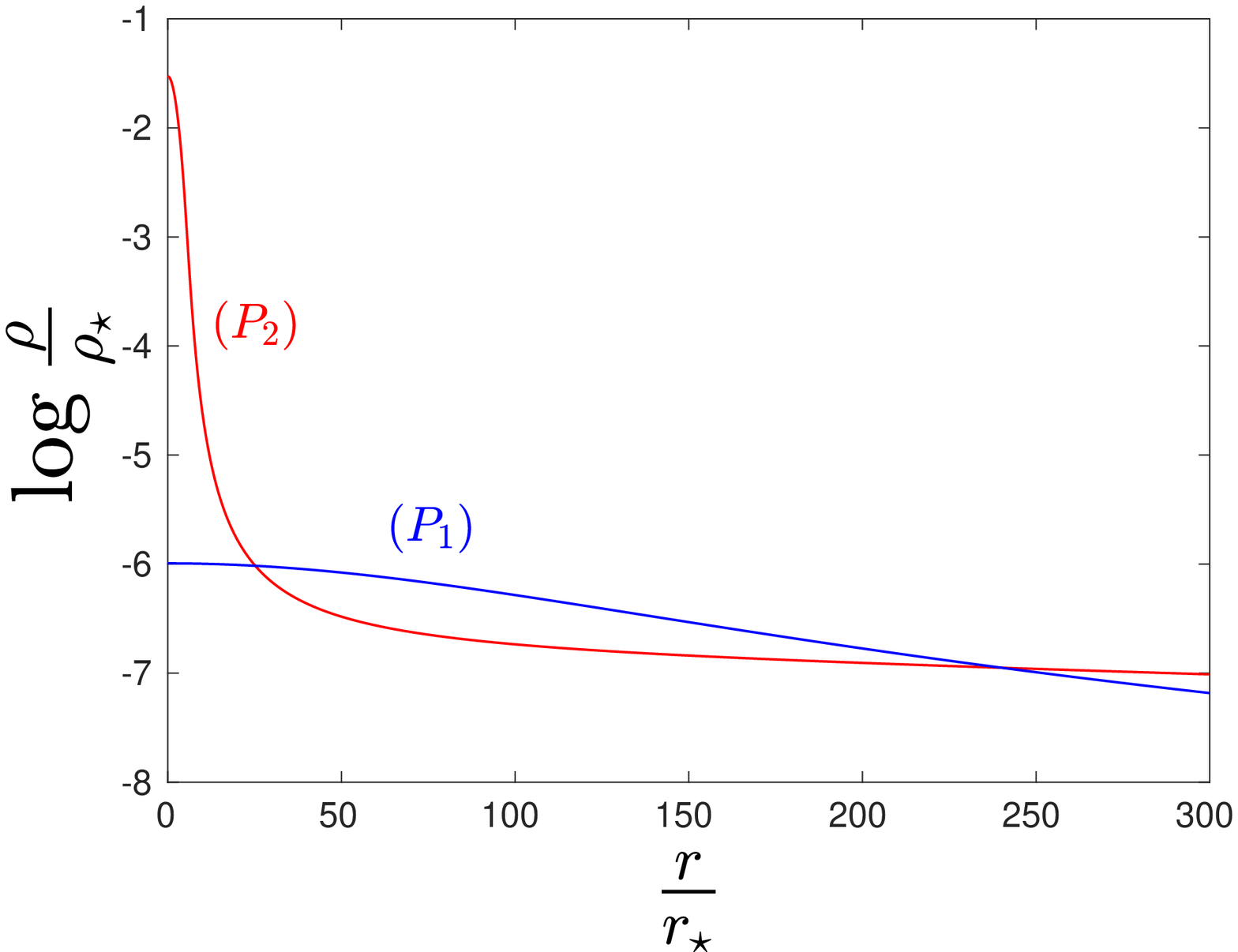} } 
	\subfigure[Temperature]{ \label{fig:T_F100_R300}
	\includegraphics[scale = 0.3]{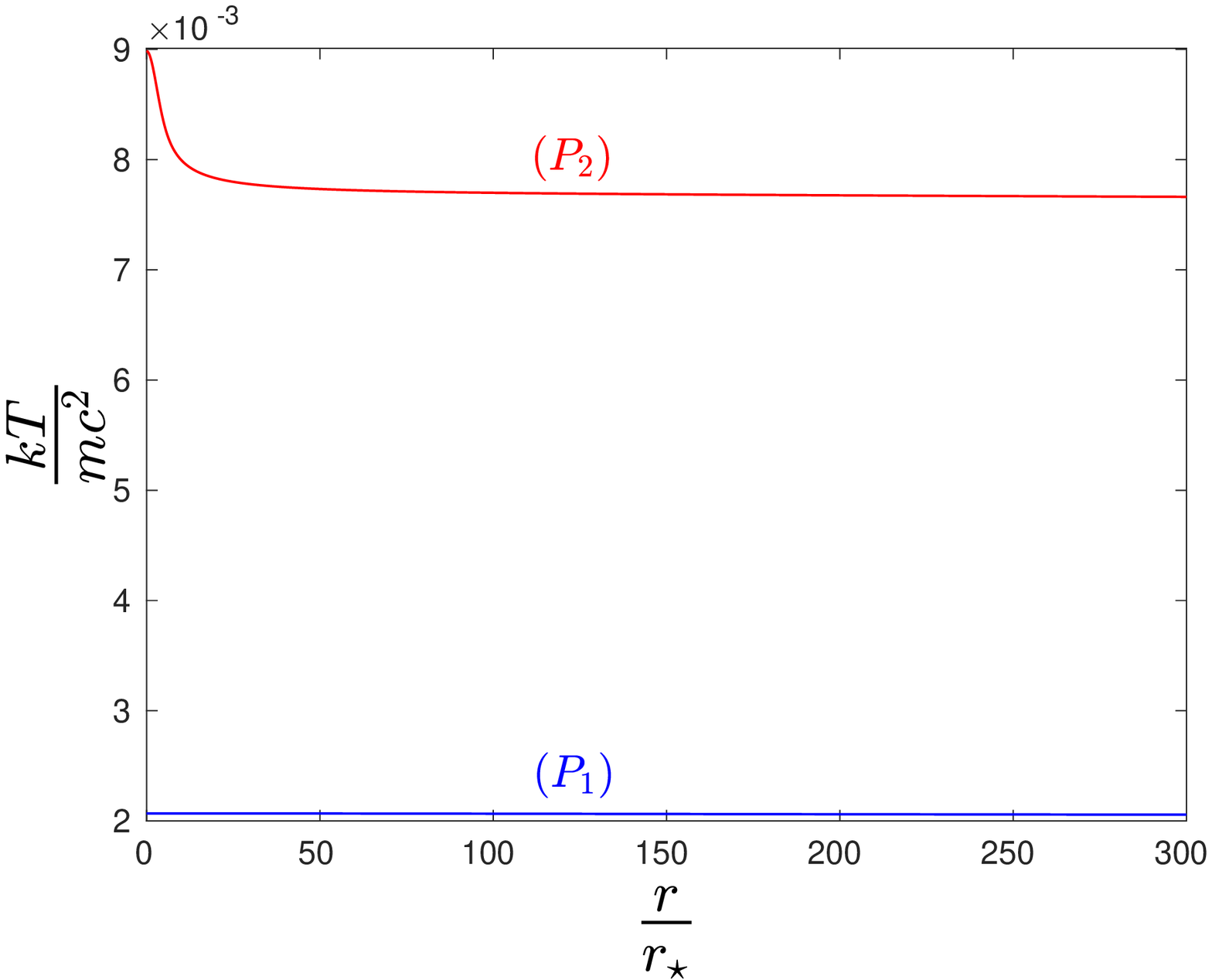} } 
	\caption{$\xi = 0.01$, $\zeta = 300$. The density and 
temperature profiles for the gaseous phase $(P_1)$, which corresponds to point
$P_1$ of Figure \ref{fig:E_F100_R300}, and the condensed phase $(P_2)$, which
corresponds to point $P_2$ of Figure \ref{fig:E_F100_R300}. Clearly $(P_2)$
consists of an ultra-dense core and a diluted halo, while $(P_1)$ is
comparatively nearly homogeneous. The condensed  phase occurs at higher
temperature than the gaseous phase and the core is substantially
hotter than the halo
with a significant Tolman-Ehrenfest effect taking place.
In contrast, the temperature of the gaseous phase is nearly homogeneous. This is because the
gaseous phase is weakly relativistic while the condensed phase is strongly relativistic. Both
configurations have common entropy corresponding to point $P$ of Figure
\ref{fig:S_F100_R300}. 
	\label{fig:rhoT_F100_R300}}
\end{center} 
\end{figure}

There appear two marginal values $\zeta_{\rm
MCP}=154$
and $\zeta_c=396$ of the system
size. For $\zeta < \zeta_{\rm MCP}=154$ the gravothermal catastrophe is
suppressed and
does not occur as in the cases $\zeta=10,80$ of Figure \ref{fig:B_E_F100}. 
For $\zeta_{\rm MCP}=154 < \zeta < \zeta_c=396$ the gravothermal catastrophe
does occur at $E_A$,
but it is
halted by a degenerate configuration, as in the case $\zeta=300$ in Figures
\ref{fig:B_E_F100} and \ref{fig:F100_R300}. In this
case a \emph{gravitational phase transition} takes place from the gaseous phase
to the
condensed phase. However, this (nearly) degenerate
configuration 
undergoes a new type of instability on its turn at
$E_C$.\footnote{For $\xi>\xi'_{\rm MCP}=0.00461$, the size $\zeta_{\rm
MCP}(\xi)$ at which the microcanonical phase transition occurs is larger than
the size $\zeta_{\rm OV}(\xi)$ at which the ground state disappears. As a
result, the condensed phase always collapses at sufficiently low energies. For
$\xi<\xi'_{\rm MCP}=0.00461$ there is an interval of sizes $\zeta_{\rm
MCP}(\xi)<\zeta<\zeta_{\rm
OV}(\xi)$ where this relativistic instability does not take place.}
This happens when the number of particles $N_C$ in the core passes above the OV limit $N_{\rm OV}$ leading to core collapse.  
The turning point of this instability is denoted by the letter $C$ in Figure \ref{fig:E_F100_R300}, while letter $A$ denotes the gravothermal catastrophe. 
The density and temperature distributions of the two phases, core-halo phase and
gaseous phase, are given in Figure \ref{fig:rhoT_F100_R300}. For $\zeta >
\zeta_c=396$ the gravothermal catastrophe not only occurs but
collapse at $E_A$
cannot be halted because $E_C>E_A$ or because every condensed
configuration is unstable as in the case
$\zeta=1000$ of Figure \ref{fig:B_E_F100} (the condensed phase disappears or
becomes unstable for $\zeta > \zeta'_*$ slightly larger than $\zeta_c$). 

\begin{figure}[htb]
\begin{center}
	\subfigure[Compactness]{ \label{fig:M_x_F100_R300_Emin}
	\includegraphics[scale = 0.4]{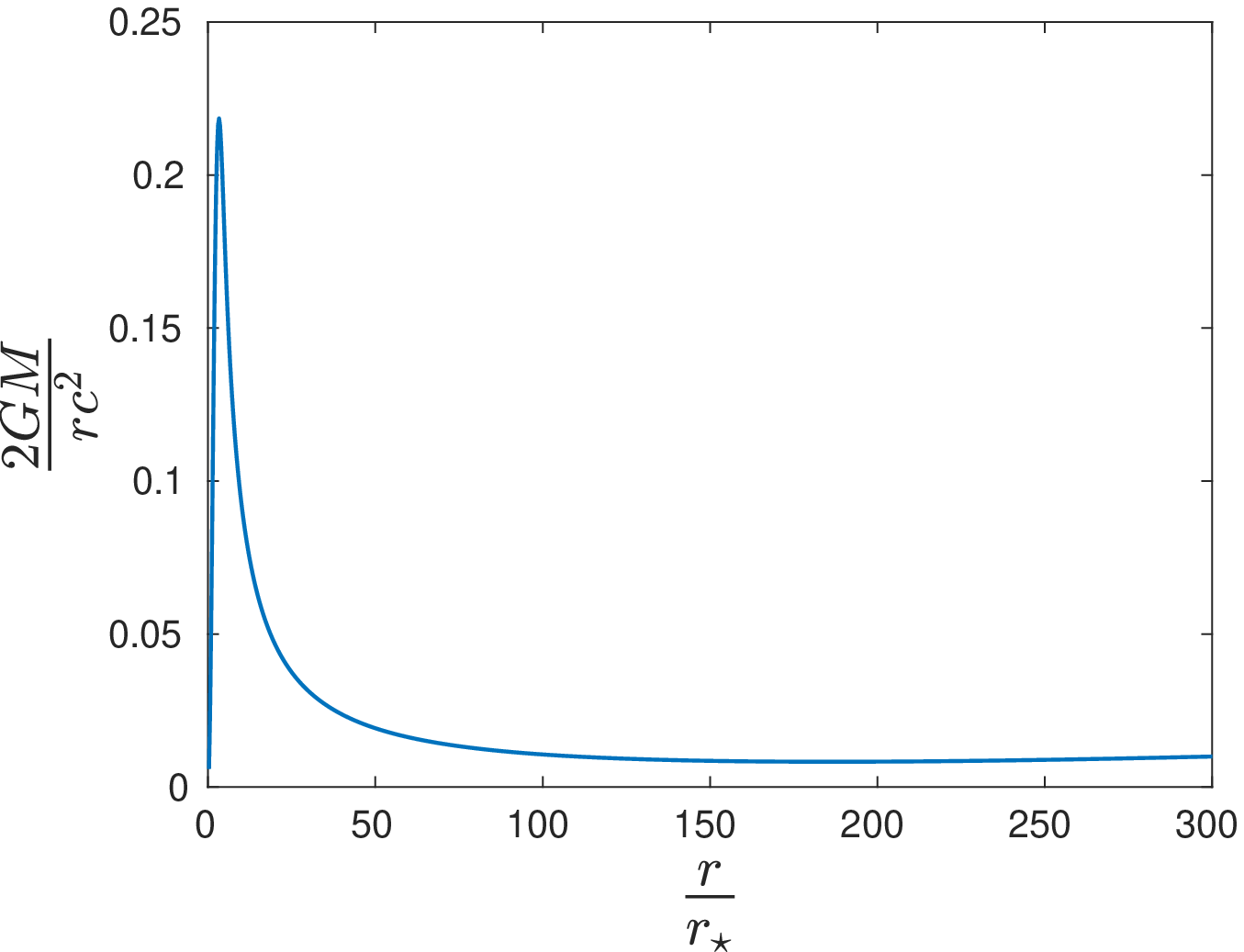} } 
	\subfigure[Temperature]{ \label{fig:T_x_F100_R300_Emin}
	\includegraphics[scale = 0.4]{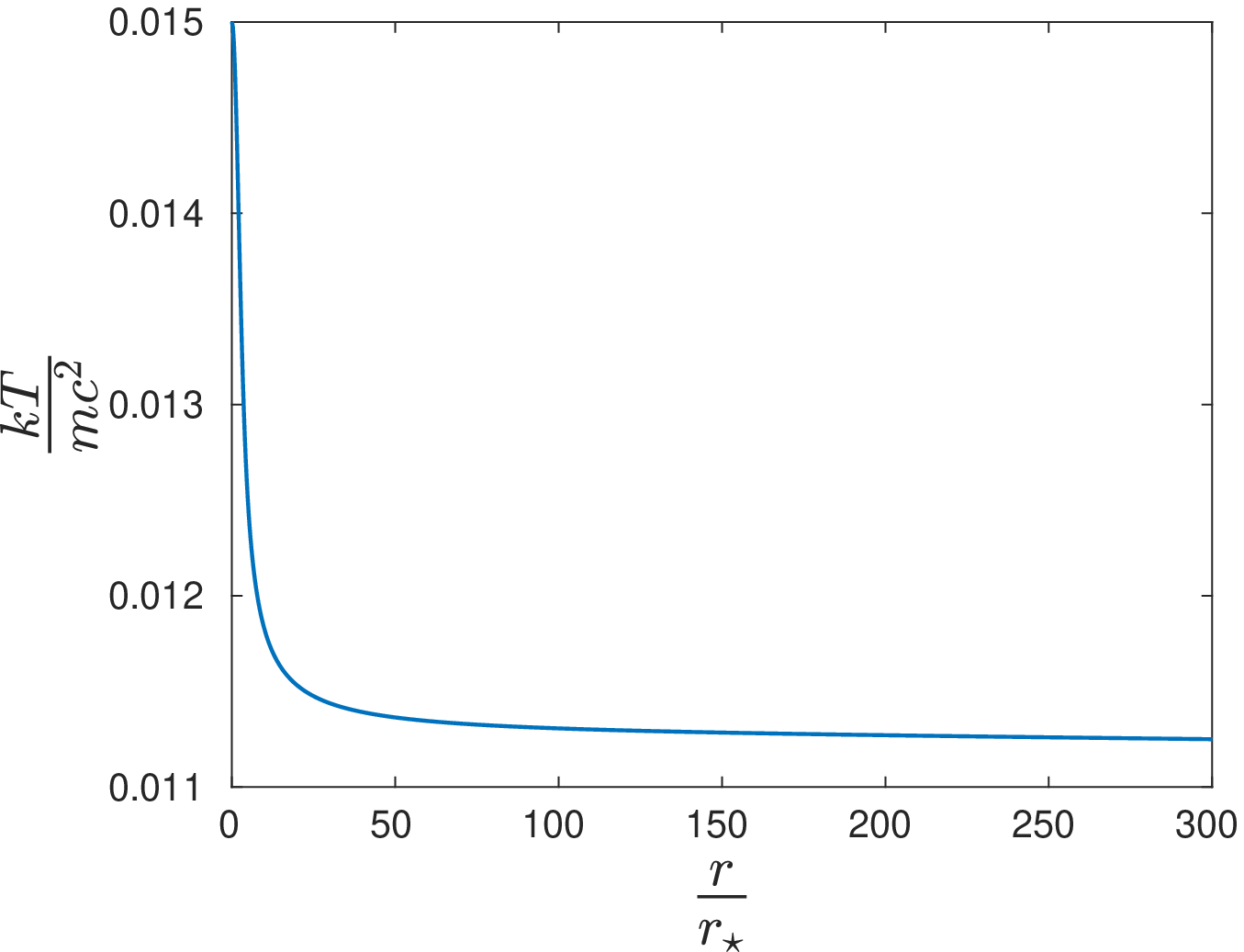} } 
	\caption{$\xi = 0.01$, $\zeta = 300$. The compactness and 
temperature profiles of the equilibrium at point $C$ of Figure
\ref{fig:E_F100_R300}, where a relativistic instability sets in. Although for
the whole sphere the compactness $2GM/Rc^2 \simeq 2G\mathcal{M}/Rc^2
= 0.01$ is small, the  Newtonian approximation is not  valid because the system
has developed a very relativistic dense core ($R_C\ll R$) for which 
$2GM_{C}/R_{C} c^2 \simeq
0.22$, leading to a general relativistic instability. 
	\label{fig:MT_F100_R300_Emin}}
\end{center} 
\end{figure}

\begin{figure}[htb]
\begin{center}
	\subfigure[Inverse temperature]{ \label{fig:B_E_F100_R80}
	\includegraphics[scale = 0.3]{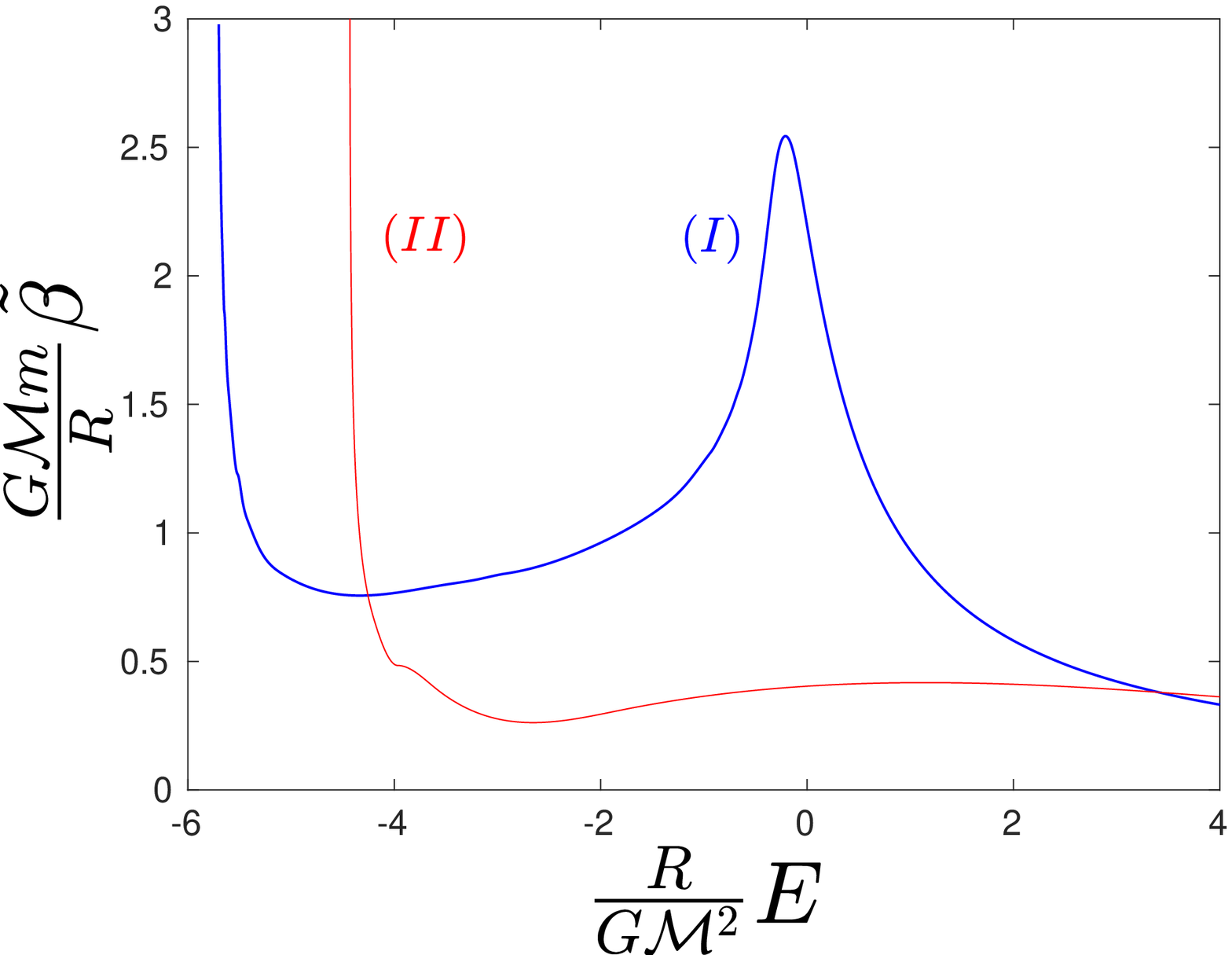} } 
	\subfigure[Entropy]{ \label{fig:S_F100_R80}
	\includegraphics[scale = 0.3]{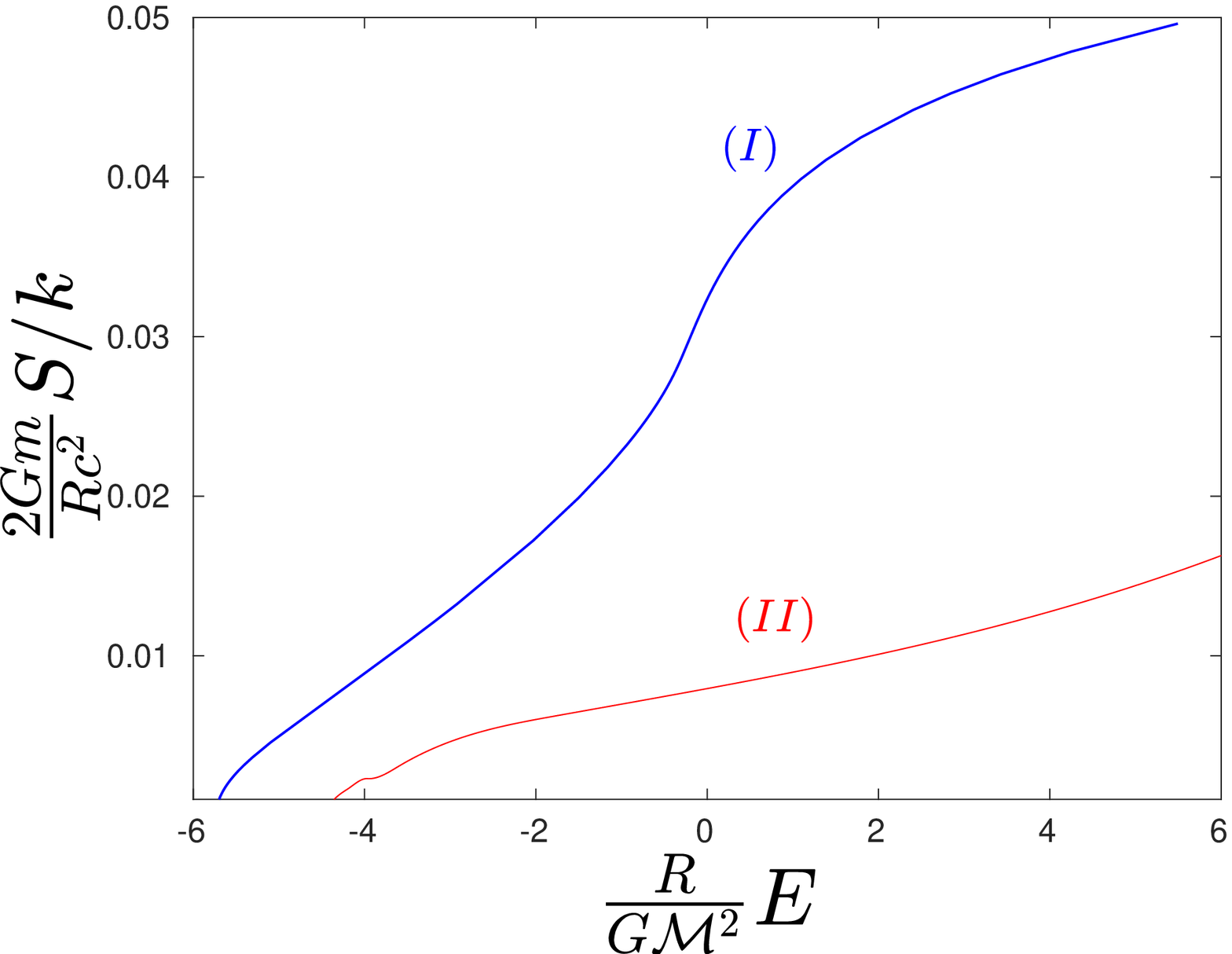} } 
	\caption{$\xi = 0.01$, $\zeta = 80$. Above  $\zeta_1=40.8$  appear
two distinct 
series of equilibria. Blue, thick branch $(I)$ is stable and has a higher
entropy than the red thin unstable branch $(II)$. It is the analogue of branch
$(IV)$ of Figure \ref{fig:E_F100_R300}, but separated. The branches present
two vertical  asymptotes at $E_{\rm min}$ and 
$E'_{\rm min}$ where $\tilde\beta\rightarrow +\infty$. They correspond to the
stable and first unstable equilibrium states of
the general relativistic Fermi gas at $T=0$. The two series of
equilibria merge at $\zeta_{\rm
OV}=90.0$ where the asymptotes disappear.
	\label{fig:F100_R80}}
\end{center} 
\end{figure}

The series of equilibria are intersecting 
in various cases as in Figures \ref{fig:E_F100_R300} and
\ref{fig:B_E_F100_R80}. This does not a raise a problem,
because what is plotted
is the Tolman temperature. So there is a third parameter
(apart
from the energy
and the Tolman temperature) that defines each configuration. This is the central
temperature $b_0$. Therefore at the point of intersection there correspond two
distinct equilibria with the same Tolman temperature and energy but with
different central temperature.  As
already implied, the
equilibrium in the condensed phase has a much larger central
temperature so that the core is much hotter than the center region of the
corresponding gaseous phase that does not posses a core. Note, however, that as we explained earlier the temperature normalized to the corresponding Fermi temperature is smaller for the core than the gaseous phase.
Only one of these two configurations is stable. The condensed configurations  on branch (IV) in
Figure \ref{fig:E_F100_R300} and on branch (II) in
Figure \ref{fig:B_E_F100_R80} are unstable while the gaseous configurations
on branch (I) in Figures \ref{fig:E_F100_R300} and \ref{fig:B_E_F100_R80} 
 are stable.

\begin{figure}[h!t]
\begin{center}
	\includegraphics[scale = 0.4]{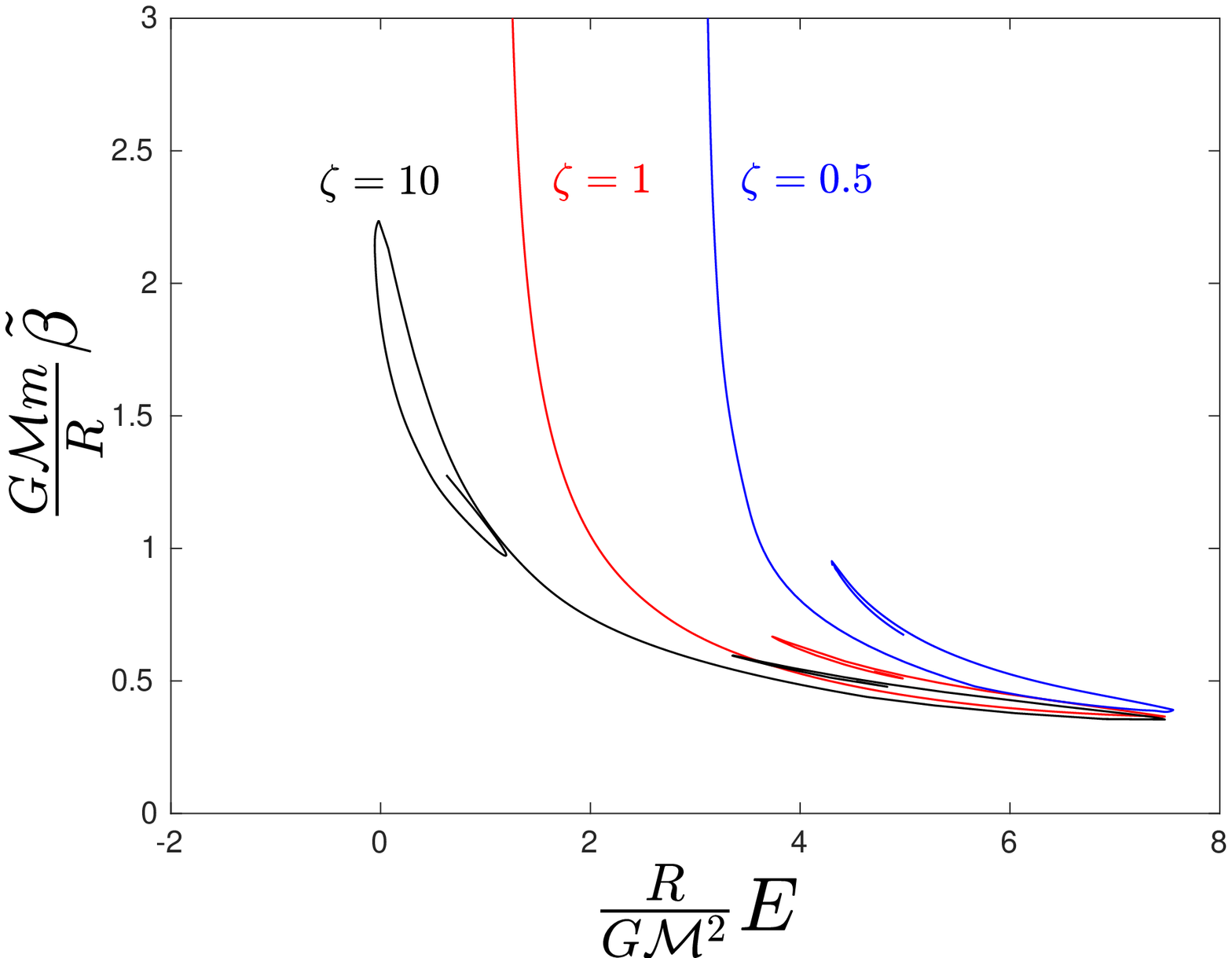}
	\caption{$\xi = 0.25$. This value of $\xi$ corresponds to a very strong
gravitational field. The phase transition to a condensed phase is
suppressed. There are only two possibilities depending on the value of the
system's size $\zeta$: the appearance of the low energy
gravothermal instability when $\zeta>\zeta_{\rm OV}=3.60$ or the complete
suppression of it and the existence of
equilibria until $\tilde{T}=0$ and $E_{\rm min}$ (ground state) when
$\zeta<\zeta_{\rm OV}=3.60$. At sufficient high energy there appears the high-energy gravothermal instability \cite{roupas,Roupas_RGI_2018} for any value of compactness and size.
	\label{fig:B_E_F4}}
\end{center} 
\end{figure}

In Figure \ref{fig:B_E_F4} are drawn the caloric curves for $\xi=0.25$ and
various $\zeta=0.5,\; 1,\; 10$ (here $\zeta_{\rm min}=0.185$). This value of
$\xi$ corresponds to a very strong gravitational field where general relativity
cannot be ignored at any case.
For this value of $\xi$ the  phase transitions are suppressed for any
$\zeta$.\footnote{The microcanonical phase
transition completely disappears above 
$\xi_{\rm MCP}=0.0272$ while the canonical phase transition completely
disappears above $\xi_{\rm
CCP}=0.0707$.} For $\zeta<\zeta_{\rm OV}=3.60$ the first branch presents an  
 asymptote at $E_{\rm min}$ where $\tilde\beta\rightarrow +\infty$. A second
branch with an asymptote at $E'_{\rm min}$  appears at
$\zeta_{1}=1.45$ (this branch is unstable, similar to branch (II)
in Figure \ref{fig:B_E_F100_R80}, and is not presented).
The first and second branches  (i.e. the asymptotes at $E_{\rm min}$ and $E'_{\rm min}$) merge at
$\zeta_{\rm OV}=3.60$.  

\section{Summary and conclusion}\label{sec:conclusions}

We have provided an
illustration of
microcanonical phase transitions
and instabilities in the general relativistic Fermi gas at nonzero temperature.
We have specified a value of the rest mass compactness $\xi=2G{\cal M}/Rc^2$ and
studied the caloric curves as a
function of the system size $\zeta=R/r_*$.
We have first considered a low value of the
rest mass compactness, $\xi=0.01$, so that our system is expected to be close
to the Newtonian
gravity limit. For $\zeta<\zeta_{\rm MCP}$ there is no phase transition
but a region of negative specific heat appears for
$\zeta>\zeta_{\rm CCP}$. For $\zeta>\zeta_{\rm MCP}$ the system undergoes a
gravothermal
catastrophe at some critical relativistic Antonov energy $E_A$. For $\zeta_{\rm
MCP}<\zeta<\zeta_c$ the gravothermal catastrophe is halted by quantum
degeneracy (Pauli's exclusion principle) so that a microcanonical phase
transition from a gaseous phase to a condensed phase occurs. However, at a lower
energy $E_C$, the condensed phase undergoes a relativistic instability which occurs for
$\zeta>\zeta_{\rm OV}$. This is because the condensed phase has a core-halo
structure and the degenerate core becomes relativistically unstable.
For $\zeta>\zeta_c$, we find that quantum mechanics cannot arrest the
gravothermal catastrophe at $E_A$ so that the gaseous phase collapses 
without passing through a condensed state. We
have then considered a higher
value of the rest mass compactness, $\xi=0.25$, corresponding to a
strongly relativistic regime. In that case, there is no phase transition.
However, a low-energy gravothermal instability occurs
for $\zeta>\zeta_{\rm OV}$. The high-energy gravothermal instability appears for any values of the control parameters $\xi$, $\zeta$. This is evidence of its universal character \cite{Roupas_RGI_2018}.

\appendix

\section{Domains of validity of the different regimes}
\label{app:sec_pd}

In this Appendix, we determine the domains of validity of the different regimes as a function of $\xi$ and
$\zeta$ using the general results from \cite{ac2}.

\subsection{The general relativistic Fermi gas at $T=0$}
\label{sec_ov}

\begin{figure}
\begin{center}
\includegraphics[clip,scale=0.3]{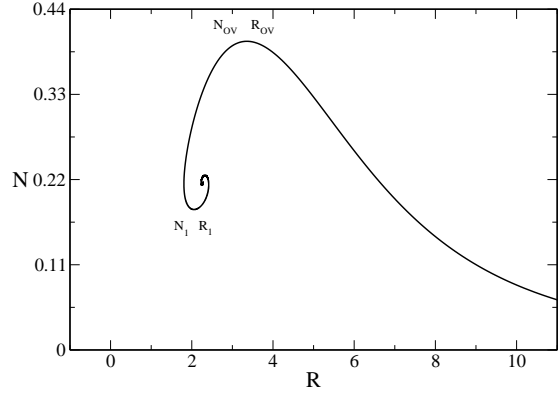}
\caption{Particle number-radius relation for the general relativistic Fermi gas
at $T=0$.}
\label{nr}
\end{center}
\end{figure}

We first consider the general relativistic Fermi gas at $T=0$. This model was
originally studied by
Oppenheimer and Volkoff \cite{ov} in the context of neutron stars. In that case,
the system is self-confined and the material box is not necessary.
The particle number-radius relation is plotted in Fig. \ref{nr}. It
has a
snail-like (spiral) structure. There is no equilibrium state above a maximum
particle number
\begin{equation}
\label{dq2}
N_{\rm OV}=0.39853\, \sqrt{\frac{2}{g}}\left (\frac{\hbar c}{G}\right
)^{3/2}\frac{1}{m^3}=  0.449693 \,    N_\star.
\end{equation}
The corresponding maximum mass and minimum radius are   
\begin{equation}
\label{dq1}
M_{\rm OV}=0.38426\, \sqrt{\frac{2}{g}}\left (\frac{\hbar c}{G}\right
)^{3/2}\frac{1}{m^2}=0.433591\, M_\star,
\end{equation}
\begin{equation}
\label{dq3}
R_{\rm OV}=8.7360\, \frac{GM_{\rm OV}}{c^2}=3.3569\,
\sqrt{\frac{2}{g}}\left
(\frac{\hbar^3}{Gc}\right
)^{1/2}\frac{1}{m^2}= 3.78786\, r_\star.
\end{equation}
For $N<N_1$, where
\begin{equation}
\label{dq2b}
N_{1}=0.18131\, \sqrt{\frac{2}{g}}\left (\frac{\hbar c}{G}\right
)^{3/2}\frac{1}{m^3}=0.20459\,    N_\star,
\end{equation}
with  corresponding mass and radius
\begin{equation}
\label{dq1b}
M_{1}=0.19893\, \sqrt{\frac{2}{g}}\left (\frac{\hbar c}{G}\right
)^{3/2}\frac{1}{m^2}=0.22447\, M_\star,
\end{equation}
\begin{equation}
\label{dq3b}
R_{1}=2.0556\,
\sqrt{\frac{2}{g}}\left
(\frac{\hbar^3}{Gc}\right
)^{1/2}\frac{1}{m^2}=2.3195\, r_\star,
\end{equation}  
there is only one equilibrium state at $T=0$ and it is stable. For $N_1<N<N_{\rm
OV}$ there are two or more  (up to an infinity) equilibrium states at $T=0$.
However, only the equilibrium states on the main branch, before the first
turning point of $N(\epsilon_0)$ at $N_{\rm max}$, are stable
(this corresponds to a mass-radius
ratio less than $(2GM/Rc^2)_{\rm max}=0.229$). The other
equilibrium states are
unstable and they have more and more modes of instability as the $N(R)$ curve
spirals inwards. Below, we shall consider only the first unstable equilibrium
state. It appears suddenly at $N_1$ (as we increase $N$) and merges with the
stable equilibrium 
state at $N_{\rm
OV}$.

This observation allows us to understand one important feature of the caloric
curves of the general relativistic Fermi gas. For
$N<N_{\rm OV}$ there exists a stable equilibrium state at $T=0$. This is the
limit point of the main branch of the caloric curve ending on an asymptote
at $E_{\rm min}$ and $\tilde\beta=+\infty$
(ground state).  For $N_1<N<N_{\rm OV}$ there exists in addition an unstable
equilibrium state at $T=0$. This is the limit
point of the secondary branch of the caloric curve ending on an asymptote at 
$E'_{\rm min}$ and $\tilde\beta=+\infty$ (at $N=N_1$ we have $E'_{\rm
min}R/G{\cal M}^2=0.53617R$ and
$E_{\rm min}R/G{\cal M}^2=-0.0570R$). The two branches merge at $N_{\rm
OV}$ where the asymptotes at $E_{\rm min}$ and $E'_{\rm min}$ meet each other
(at $N=N_{\rm OV}$ we have $E'_{\rm
min}R/G{\cal M}^2=E_{\rm min}R/G{\cal M}^2=-0.08985R$). For
$N>N_{\rm OV}$ there is no ground state, i.e., there is no stable equilibrium
state
at $T=0$ anymore. As a result, there is no vertical asymptote in the caloric
curve. In that case, the system undergoes gravitational collapse
below a critical temperature or below a critical energy. They correspond to
turning points of temperature $\tilde T(\epsilon_0)$ and energy $E(\epsilon_0)$
in the series of equilibria.

\subsection{The phase diagram of the nonrelativistic Fermi gas}
\label{sec_nfg}

In the nonrelativistic limit, the caloric curve of the
self-gravitating Fermi gas
depends on a single control parameter (instead of depending on $N$ and $R$
individually) which
can be written as \cite{ijmpb}:
\begin{equation}
\label{nfg1}
\mu =\sqrt{2\frac{N}{N_{\star}}\frac{R^3}{r_\star^3}}.
\end{equation}
The phase diagram of the nonrelativistic self-gravitating Fermi gas is given in
\cite{ijmpb}. It is shown in this paper that a canonical
phase transition appears above $\mu_{\rm CCP}=83$ and that a
microcanonical phase transition appears above $\mu_{\rm MCP}=2670$. For a given
radius $R$, using equation (\ref{nfg1}), we conclude that the canonical
phase transition appears above the  particle number
\begin{equation}
\frac{N^{\rm NR}_{\rm
CCP}(R)}{N_*}= 3.44\times 10^3 \left (\frac{r_*}{R}\right )^3,
\end{equation}
and that the  microcanonical
phase transition appears above the  particle number
\begin{equation}
\frac{N^{\rm NR}_{\rm
MCP}(R)}{N_*}= 3.57\times 10^6 \left (\frac{r_*}{R}\right )^3.
\end{equation}

\subsection{The phase diagram of the general relativistic Fermi gas in the
$(R,N)$ plane}
\label{app:sec_ac}

\begin{figure}
\begin{center}
\includegraphics[clip,scale=0.3]{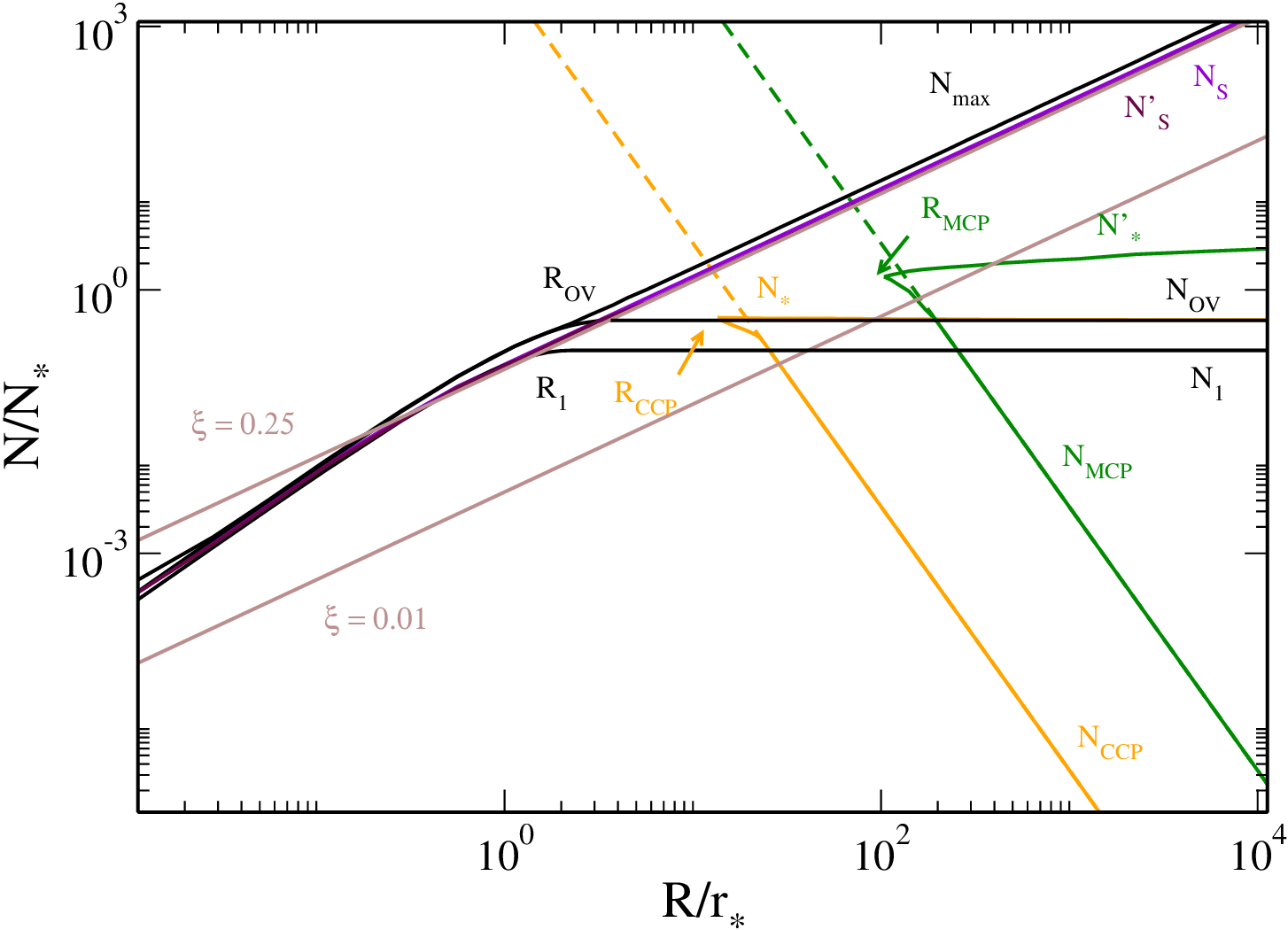}
\caption{The phase diagram of the general relativistic Fermi gas in the
$(R,N)$ plane (taken from
\cite{ac2}).}
\label{diakphaseNORMALISATIONROUPAS}
\end{center}
\end{figure}

In the general relativistic case, the caloric curves of the
self-gravitating Fermi gas depend on $R$ and $N$ individually. The
phase diagram of the general relativistic Fermi gas in the $(R,N)$ plane
has been obtained in \cite{ac2}. It is reproduced in
Figure \ref{diakphaseNORMALISATIONROUPAS} with the notations of the present
paper. We  recall below the meaning of the different curves (we refer to
\cite{ac2} for a more detailed description): 

(i) The curve $N_{1}(R)$ signals the appearance of  a second branch of
solutions in the caloric curve (corresponding to unstable equilibrium states).
For $R/r_*>R_1/r_*=2.32$, we have $N_{1}/N_*=0.204$.
For
$R/r_*\ll R_1/r_*=2.32$, we have $N_{1}(R)/N_*\sim
0.234 (R/r_*)^{3/2}$.\footnote{This change of regime, here and in points (ii)
and (iii) below, is due to the fact that the self-gravitating Fermi gas at
$T=0$ is
confined by the box, instead of being self-confined, when the box radius is too
small.}

(ii) The curve $N_{\rm OV}(R)$ signals the disappearance of the ground state
(i.e. there is no equilibrium state at $T=0$ anymore). At that point, the
asymptotes at $E_{\rm min}$ and $E'_{\rm min}$ of the first and
second branches in the caloric curve merge, then disappear. For $R/r_*>R_{\rm
OV}/r_*=3.79$, we have
$N_{\rm OV}/N_*=0.450$. For $R/r_*\ll R_{\rm
OV}/r_*=3.79$, we
have $N_{\rm OV}(R)/N_*\sim
0.292 (R/r_*)^{3/2}$. 

(iii)  The curve $N_{\rm max}(R)$ is the maximum particle number for
which there
are equilibrium states.  For $R/r_*\gg R_{\rm OV}/r_*=3.79$, we have
$N_{\rm max}(R)/N_*=0.1764 R/r_*$. For $R/r_*\ll R_{\rm
OV}/r_*=3.79$, we
have $N_{\rm max}(R)/N_*\sim
0.292 (R/r_*)^{3/2}$.

(iv) There is no canonical phase transition when $R/r_*< R_{\rm CCP}/r_*=13.5$.
When $R/r_*> R_{\rm CCP}/r_*=13.5$, the curve $N_{\rm CCP}(R)$ signals the
appearance of a canonical phase
transition. For $R/r_*\gg R_{\rm CCP}/r_*=13.5$, we have $N_{\rm
CCP}/N_*\sim 3.44\times 10^3 (r_*/R)^3$. The curve
$N_{*}(R)$ signals the
disappearance of the condensed phase in the canonical
ensemble. Note that $N_{*}(R)$ is very close to the value
$N_c^{\rm CE}(R)$ at which the isothermal collapse is not halted by quantum
mechanics.

(v) There is no microcanonical phase transition when $R/r_*< R_{\rm
MCP}/r_*=104$. When $R/r_*> R_{\rm
MCP}/r_*=104$, the curve $N_{\rm MCP}(R)$ signals the appearance of
a microcanonical phase
transition. For $R/r_*\gg R_{\rm MCP}/r_*=104$, we have $N_{\rm
MCP}/N_*\sim 3.57\times 10^6 (r_*/R)^3$. The
curve $N_{*}'(R)$ signals the disappearance of the condensed phase in the 
microcanonical ensemble. Note that $N_{*}'(R)$ is very close to
the value
$N_c^{\rm MCE}(R)$ at which the gravothermal catastrophe is not halted by
quantum
mechanics.

\subsection{The $(\xi,\zeta)$ variables}
\label{sec_ac}

In the present paper, we have taken the rest mass compactness
\begin{equation}
\xi = \frac{2GNm}{Rc^2}=\frac{2N/N_\star}{R/r_\star} 
\end{equation}
and the box radius
\begin{equation}
\zeta = \frac{R}{r_\star}
\end{equation}
as control parameters. We shall fix the relativistic parameter
$\xi$ and describe the caloric curves and the phase transitions as a function of
the box radius $\zeta$, using the phase diagram of Fig.
\ref{diakphaseNORMALISATIONROUPAS}. We note that fixing $\xi$ determines a
straight line of equation $N/N_\star=(\xi/2)(R/r_\star)$ in the phase diagram of
Fig. \ref{diakphaseNORMALISATIONROUPAS}. Therefore, changing $\zeta$ at fixed
$\xi$ amounts to moving along that line. For a fixed value of $\xi$, we find
that:

(i) There is no equilibrium state above $\xi_{\rm max}=0.353$, whatever the
value of the energy and of the temperature. 

(ii) The smallest possible value of $\zeta$ is $\zeta_{\rm min}(\xi)$. For
$\xi\ll\xi_{\rm max}=0.353$ we have $\zeta_{\rm min}(\xi)\sim 2.93\xi^2$.

(ii) The second branch in the caloric curve appears at $\zeta_1(\xi)$.  For
$\xi\ll\xi_{\rm max}=0.353$ we have  $\zeta_1(\xi)\sim 0.408/\xi$.

(iii) The ground state (equilibrium state at $T=0$) disappears at  $\zeta_{\rm
OV}(\xi)$. At that point, the asymptotes at $E_{\rm min}$ and  $E'_{\rm min}$ of
the first and second branches merge, then disappear.  For
$\xi\ll\xi_{\rm max}=0.353$ we have  $\zeta_{\rm OV}(\xi)\sim 0.900/\xi$.

(iv) There is no canonical phase transition when $\xi>\xi_{\rm
CCP}=0.0707$. When  $\xi< \xi_{\rm
CCP}=0.0707$, the canonical phase transition
appears at $\zeta_{\rm
CCP}(\xi)$. When $\xi\ll \xi_{\rm
CCP}=0.0707$, we have $\zeta^{\rm NR}_{\rm
CCP}(\xi)=9.11/\xi^{1/4}$. The  canonical phase transition disappears at 
$\zeta_{*}(\xi)$. Note that
$\zeta_{*}(\xi)$ is very close to the value $\zeta_c^{\rm CE}(\xi)$ at which the
isothermal collapse is not halted by quantum mechanics.

(v)  There is no microcanonical phase transition when $\xi>\xi_{\rm
MCP}=0.0272$. When $\xi< \xi_{\rm
MCP}=0.0272$,  the microcanonical phase transition
appears at $\zeta_{\rm MCP}(\xi)$. When $\xi\ll \xi_{\rm
MCP}=0.0272$, we have $\zeta^{\rm NR}_{\rm
MCP}(\xi)=51.7/\xi^{1/4}$. The  microcanonical phase transition disappears at 
$\zeta'_{*}(\xi)$. Note that
$\zeta'_{*}(\xi)$ is very close to the value $\zeta_c^{\rm MCE}(\xi)$ at which
the
gravothermal catastrophe is not halted by quantum mechanics. Two situations
may occur. Let us first assume 
$\xi<\xi'_{\rm MCP}=0.00461$. In that case: when $\zeta_{\rm
MCP}(\xi)<\zeta<\zeta_{\rm
OV}(\xi)$ the condensed phase is stable for all energies because $N<N_{\rm OV}$;
when $\zeta_{\rm OV}(\xi)<\zeta<\zeta'_{*}(\xi)$ the condensed phase collapses
at small energies  because $N>N_{\rm OV}$.  Let us now assume $\xi'_{\rm
MCP}=0.00461<\xi<\xi_{\rm MCP}=0.0272$. In that case,  the condensed phase
collapses at small energies  because $N>N_{\rm OV}$.

In this paper, for illustration, we have considered  two specific values of
$\xi$. 

For $\xi=0.01$, we plot on Fig. \ref{diakphaseNORMALISATIONROUPAS} the straight
line $N/N_\star=0.005 R/r_\star$. We have  $\zeta_{\rm
min}=2.93\times 10^{-4}$.  The second
branch appears at $\zeta_{1}=40.8$. The
first and second branches merge at $\zeta_{\rm
OV}=90.0$. The canonical phase
transition appears at $\zeta_{\rm
CCP}=28.7$ and ends at $\zeta_{*}=93.5$. The
microcanonical phase
transition
appears at $\zeta_{\rm MCP}=154$ and ends at
$\zeta'_{*}=395.7$.

For $\xi=0.25$,   we plot on Fig. \ref{diakphaseNORMALISATIONROUPAS} the
straight line $N/N_\star=0.125
R/r_\star$. We have  $\zeta_{\rm
min}=0.185$. The
second
branch appears for $\zeta_{1}=1.45$. The
first and second branches merge at $\zeta_{\rm
OV}=3.60$. There is no canonical and
no microcanonical phase transition.

\subsection{Validity of the nonrelativistic and classical limits}
\label{sec_val}

As discussed in detail in \cite{ac2}, the nonrelativistic limit corresponds to
$R\rightarrow +\infty$ and $N\rightarrow 0$ (physically $R\gg R_{\rm OV}$ and
$N\ll N_{\rm OV}$) with $NR^3$ fixed. This corresponds to the lower right panel
of Fig. \ref{diakphaseNORMALISATIONROUPAS}. In terms of the variables
($\xi,\zeta$), for a given value of $\xi\ll \xi_{\rm max}$, this corresponds to
$1\ll \zeta\ll \zeta_{\rm OV}(\xi)$ and $\zeta\gg\zeta_c(\xi)$ (these two
distinct regions are explained in Sec. XI of \cite{ac2}). On the other hand,
the classical limit
corresponds to $R\rightarrow +\infty$ and $N\rightarrow +\infty$
(physically $R\gg
R_{\rm OV}$ and $N\gg N_{\rm OV}$) with $N/R$ fixed. This corresponds to the
upper right panel of Fig.
\ref{diakphaseNORMALISATIONROUPAS}. In
terms of the variables ($\xi,\zeta$),   for a given value of $\xi< \xi_{\rm
max}$, this corresponds to $\zeta\gg \zeta_{\rm
OV}(\xi)$.

\section*{References}

\bibliography{Roupas-Chavanis_relativistic-fermionic-phase-transitions_2018}
\bibliographystyle{unsrt}

\end{document}